\documentclass[prb,aps,amsmath,amssymb,reprint,showpacs,floatfix,superscriptaddress,longbibliography]{revtex4-2}
\usepackage{graphicx}
\usepackage{dcolumn}
\usepackage{bm}
\usepackage{mathtools}
\usepackage{placeins}
\usepackage{enumitem}
\usepackage{xcolor}

\usepackage[normalem]{ulem}
\allowdisplaybreaks[4]

\begin{document}



\title{Transport in magnetic-topological-insulator nanoribbons containing multiple superconductor-proximitized sectors
} 




\author{Javier Osca}
\email{javier.osca@uib.cat}
\affiliation{Institute for Cross-Disciplinary Physics and Complex Systems IFISC (CSIC-UIB), E-07122 Palma, Spain} 
\affiliation{Department of Physics, University of the Balearic Islands, E-07122 Palma, Spain}
\author{Sungguen Ryu}
\affiliation{Institute for Cross-Disciplinary Physics and Complex Systems IFISC (CSIC-UIB), E-07122 Palma, Spain} 
\affiliation{Department of Physics, University of the Balearic Islands, E-07122 Palma, Spain}
\author{Rosa L\'opez}
\affiliation{Institute for Cross-Disciplinary Physics and Complex Systems IFISC (CSIC-UIB), E-07122 Palma, Spain} 
\affiliation{Department of Physics, University of the Balearic Islands, E-07122 Palma, Spain}
\author{Llorenç Serra}
\affiliation{Institute for Cross-Disciplinary Physics and Complex Systems IFISC (CSIC-UIB), E-07122 Palma, Spain} 
\affiliation{Department of Physics, University of the Balearic Islands, E-07122 Palma, Spain}

\date{January 23, 2026}

\begin{abstract}

Transport in devices with multiple proximitized sectors depends heavily on the complex phases of the pairing gaps of those sectors.
We investigate magnetic topological insulator nanoribbons in two- and three-terminal setups with proximitized sectors and asymptotic normal leads.
Our focus is on the regime of single chiral Majoranas.
The characteristic electric and thermal interferometries of chiral Majoranas can be controlled by the complex phases of the pairing.
We predict an AC Majorana effect, in which phase dynamics induced by a voltage bias generate measurable time-dependent conductance oscillations.
A three-terminal junction with superconducting islands can be used as a Majorana router when the relative complex phases are configured.
\end{abstract}


\maketitle 

\section{Introduction}
\label{sec1}

The interplay between superconductivity and topological phases of matter 
provides remarkable scenarios to engineer and manipulate exotic quasiparticle
states such as chiral Majorana modes \cite{Fu2008,Qi2011,Sato2016,Sato2017,He2019,Rachel,Lao,Liang}. 
Magnetic topological insulators (MTIs) proximitized by conventional superconductors are particularly 
promising platforms. Theoretical proposals have demonstrated that a 
quantum anomalous Hall insulator (QAHI) coupled to an s-wave superconductor 
can enter a topological superconducting phase supporting a single chiral Majorana mode 
\cite{Qi2010,Chang2013,Wang2015,Lian2018}.
In these platforms QAHI edge states can be coherently 
converted into chiral Majorana excitations at the normal–superconductor 
interfaces.

Although there are experimental hints supporting the presence of chiral Majorana modes \cite{Shen}, 
conclusive experimental evidence of chiral Majoranas is still lacking. 
In two-terminal devices, electrical conductance quantization of $0.5e^2/h$ 
due to coherent chiral Majorana transmission was predicted
\cite{Wang2015}. 
However, this result can also be attributed to direct terminal-to-terminal coupling through the parent superconductor where the QAHI edge channels can equilibrate and lose quantum coherence \cite{Kay2020,Uday2025,Ji2018,Huang2018}.
Therefore, it is worth finding 
additional theory predictions of the chiral Majorana scenario
that may help in their detection. Thermal transport 
has been proposed as a way to search for signatures of Majorana zero modes (MZMs) \cite{Klees2024}
while we have suggested interferometry approaches. In particular, conductance oscillations
in smaller devices \cite{Osca2018}, as well as the use of cutoff electrical gates to alter the 
traveled distance of the chiral Majoranas \cite{Osca2026}.

Superconductivity is characterized by a complex order parameter, 
the so-called pairing gap. In Bogoliubov-de Gennes theory, in lowest 
approximation, this is usually described by a constant gap parameter $\Delta e^{i\phi}$, with $\Delta$ real and $\phi$ a phase.
This phase can be ignored in systems with a single superconductor, since it can be absorbed in a gauge choice.
However, in cases with two or more superconductors, phase differences can lead to
remarkable physical consequences. Indeed, in Josephson junctions a supercurrent can 
propagate from one superconducting terminal to another in abscence of any potential bias,
solely due to their phase difference \cite{BaPa1982,Tink2004,Golu2004,Anto2025,Zazu2017}, or even
alter the topological phase diagram in certain devices \cite{Schiela2024}. 

In this work, we investigate how superconducting phase differences modify transport in QAHI devices containing two or three proximitized regions. Phase control of the superconducting regions allows coherent manipulation of the 
chiral Majorana paths, yielding conspicuous
resonances in the electric and heat conductances.
We developed an analytical scattering matrix approach, where each
proximitized region is treated as a chiral Majorana scatterer.
The phase dependence of the scattering matrix is fully analytical, allowing sequential 
structures to be described trough successive matrix compositions.

A remarkable consequence of phase-controlled Majorana transport is the emergence of a dynamical response driven by time-dependent superconducting phase differences. In particular, when a voltage bias is applied between superconducting regions, the phase difference evolves in time according to the Josephson relation, leading to a modulation of the interference pattern. In this work, we show that this mechanism gives rise to an AC Majorana effect, in which the conductance of the device oscillates in time as a direct consequence of chiral Majorana interference. Unlike the conventional AC Josephson effect, which manifests as a supercurrent carried by Cooper pairs, this effect is rooted in quasiparticle transport and provides a distinct, experimentally accessible signature of Majorana modes in proximitized magnetic topological insulators. Furthermore, extending the analysis to a three-terminal junction, we demonstrate that such phase control enables a Majorana router, in which the propagation path of chiral Majorana modes can be selected through an appropriate choice of superconducting phases.

This paper is organized as follows. Section \ref{sec2} presents the theoretical framework and analyzes transport in two-terminal MTI junctions containing two superconducting islands. The analytical scattering matrix is employed to derive the electrical and thermal conductances, with particular emphasis on phase-dependent interference effects. We also discuss the role of the mediating modes and their charge properties. 
We further demonstrate that time-dependent superconducting phase differences give rise to an AC Majorana effect, manifested as conductance oscillations driven by phase dynamics.
In Sec.\ \ref{sec3}, we extend the analysis to three-terminal geometries and demonstrate how phase control enables directional transport, effectively realizing a Majorana router. Finally, in Sec.\ \ref{sec4}, we summarize our main results and discuss their implications for the detection and manipulation of chiral Majorana modes in proximitized magnetic topological insulators.

\section{ Transport in Two-Terminal Proximitized MTI Junctions}
\label{sec2}

In this section, we introduce the theoretical framework used to describe transport in a two-terminal magnetic topological insulator (MTI) device with proximitized superconducting regions. We consider a nanoribbon geometry supporting quantum anomalous Hall edge states, where superconducting segments with different complex phases are patterned along the propagation direction. The superconductors inducing the proximitized islands are supposed to be grounded. Depending on the applied biases to the the leads, a current of Cooper pairs may leak through these grounded sectors \cite{Wang2015,Daniele}. We focus on the regime in which transport is mediated by a single chiral Majorana mode and analyze how the device geometry and phase configuration determine the resulting scattering processes. To this end, we employ an analytical scattering-matrix approach that relates incoming and outgoing quasiparticle amplitudes and provides direct access to electrical and thermal transport coefficients.

\subsection{Hamiltonian}

\begin{figure}[t]
  \centering
  \includegraphics[width=0.4\textwidth]{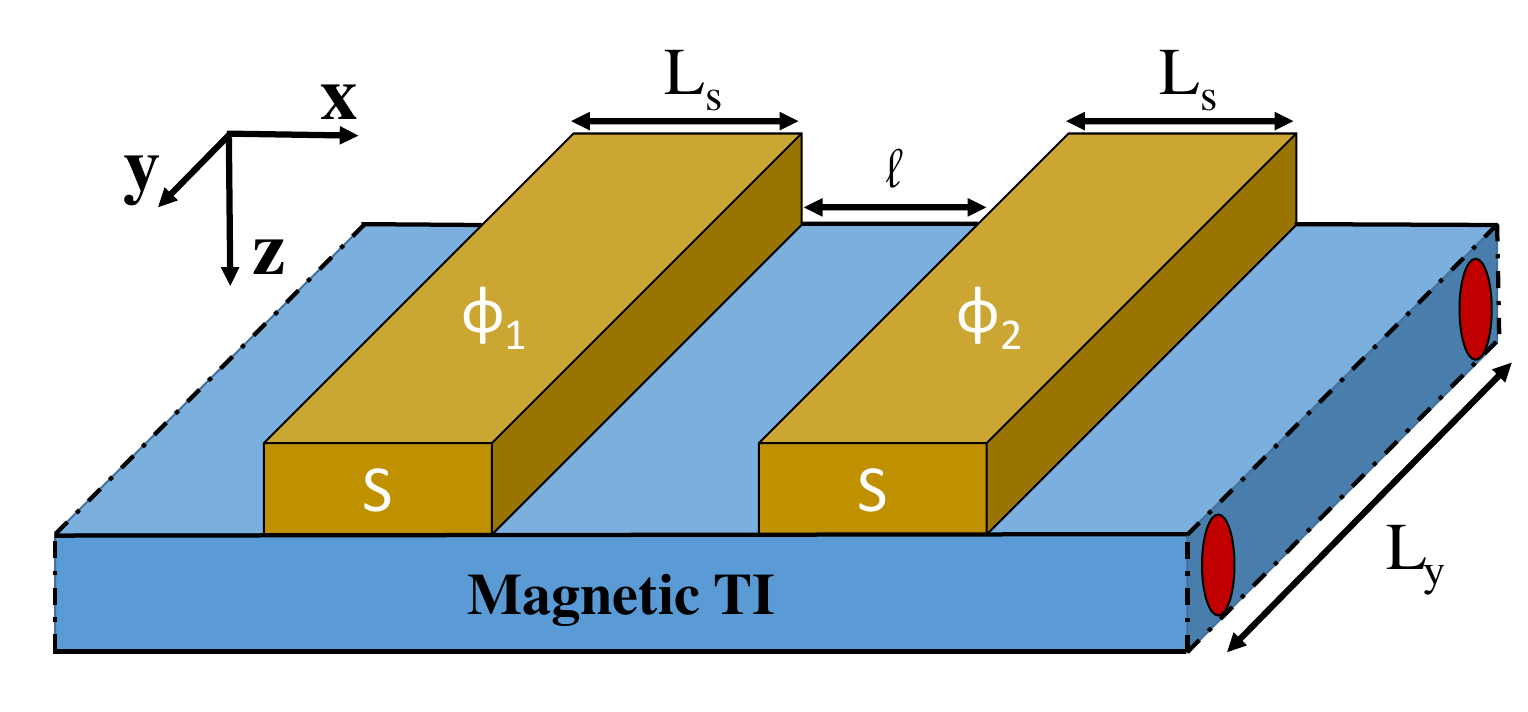}
  \caption{
Schematic of an MTI thin slab (blue) with two proximitized superconducting regions (S, shown in yellow) characterized by complex phases $\phi_1$ and $\phi_2$. Red ellipses indicate edge modes propagating along the device edges (x-direction). The figure also indicates the transverse width of the superconducting regions ($L_y$), as well as the lengths of the proximitized segments ($L_S$) and their separation ($\ell$).}
  \label{F0}
\end{figure}

We consider a MTI nanoribbon supporting QAHI edge states. The device contains two proximity-induced superconducting regions characterized by distinct complex phases, $\phi_1$ and $\phi_2$ (see Fig.~\ref{F0}). The system is described by a bilayer Hamiltonian 
\begin{eqnarray}
\label{EQ1}
{\cal H} &=& 
\left[\, m_0 + m_1 \left(p_x^2 +p_y^2\right)\, \right] \lambda_x\,\tau_z  
\nonumber\\
&-& \frac{\alpha}{\hbar}\, \left(\,p_x\sigma_y-p_y\sigma_x\right)\, 
\lambda_z \,\tau_z
+ \Delta_Z\, \sigma_z
\nonumber\\
&+& \left( \Delta_p(x) + 
\Delta_m(x)\,\lambda_z\right)e^{+i\phi(x)}  \tau_+\,\nonumber\\
&+& \left( \Delta_p(x) + 
\Delta_m(x)\,\lambda_z\right)e^{-i\phi(x)}  \tau_-\,\; .
\end{eqnarray}

In Eq.~(\ref{EQ1}), the superconducting parameters $\Delta_{p,m}(x)$ are linear combinations of the pairing amplitudes in the two layers, while $\tau_{xyz}$ and $\lambda_{xyz}$ denote Pauli matrices acting in electron–hole and layer spaces, respectively, with $\tau_{\pm}$ the corresponding ladder operators. Specifically, $\Delta_{p,m}(x) \equiv (\Delta_t \pm \Delta_b)/2$, where $\Delta_t(x)$ and $\Delta_b(x)$ are the pairing potentials in the top and bottom layers. These pairings are taken to be different, reflecting the stronger proximity effect induced by the superconductor on the top layer. Such asymmetry is required to stabilize the phase supporting a single chiral Majorana mode \cite{Wang2015}.

We model $\Delta_t(x)$, $\Delta_b(x)$, and $\phi(x)$ as piecewise constant, position-dependent functions to describe the alternation between proximitized and non-proximitized regions along the device. This spatial dependence also captures the variation of the superconducting phase between different superconducting segments. In this section, we consider a two-terminal NSNSN junction, where N and S denote normal and superconducting regions, respectively. In this context, “normal” refers to non-superconducting regions that nevertheless support topological QAH edge states.
Additionally, $\sigma_{xyz}$ denote Pauli matrices acting in spin space. The remaining parameters in Eq.~(\ref{EQ1}) are $m_0$ and $m_1$, both of which describe the interlayer coupling, $\alpha$, the strength of the Rashba-like spin–orbit interaction, and $\Delta_Z$, the Zeeman-like magnetization.

\begin{figure}[t]
\centering
\includegraphics[width=0.5\textwidth]{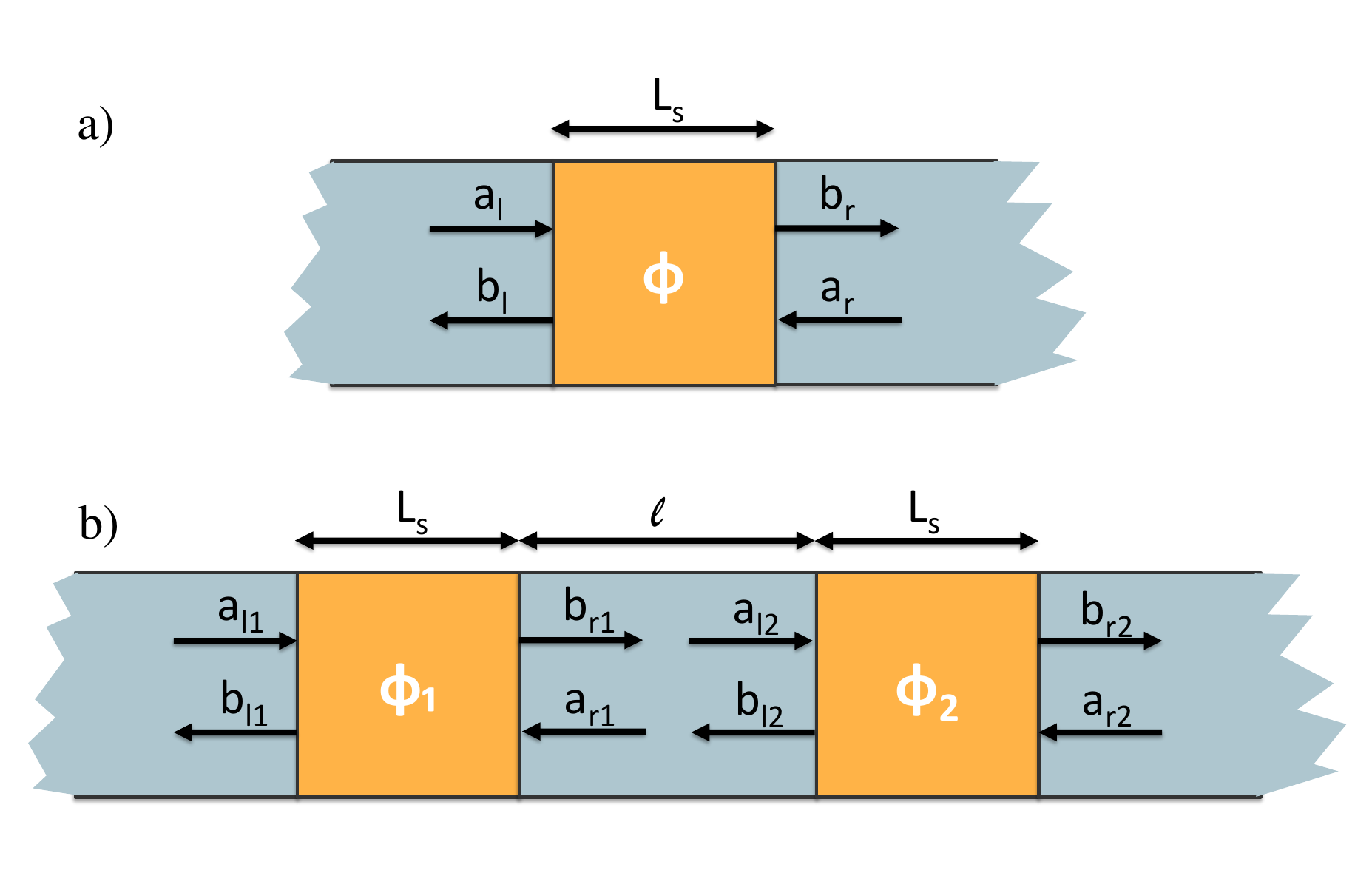}
\caption{
(a) Diagram of the scattering amplitudes in an NSN junction. (b) Same as (a) for a two-terminal NSNSN junction.}
  \label{F0B}
\end{figure}

The full NSNSN device is modeled as five distinct regions, where the two outer normal segments extend to infinity and act as contacts. The central normal region connects the two proximitized superconducting segments. These superconducting regions behave as topological superconductors and can realize different phases depending on the parameters of Eq.~(\ref{EQ1}), characterized by the topological invariant $\mathcal{N}=0,1,2$. The $\mathcal{N}=1$ phase hosts a single chiral Majorana mode at each edge, whereas the $\mathcal{N}=2$ phase is equivalent to a QAH state. From the transport perspective, the latter effectively behaves as a normal region, i.e., as if superconductivity were absent.

Since transport mediated by superconductors in the $\mathcal{N}=0$ and $\mathcal{N}=2$ phases is trivial, we restrict our analysis to the case where the superconducting segments are in the $\mathcal{N}=1$ phase. In this regime, we model each proximitized region as supporting a pair of chiral Majorana modes attached to opposite edges, propagating in opposite directions (left on the lower edge and right on the upper edge). At the interface between a QAH region and a topological superconductor, electron and hole quasiparticle states are converted into chiral Majorana modes, denoted by $\gamma_1$ and $\gamma_2$.

In general, a gauge-invariant description of a quantum state is defined only up to a global phase factor. In the present case, the electron and hole modes must therefore be written explicitly to include a phase dependence on the superconducting order parameter $\phi$. Below it will be shown how this phase choice reflects the phase accumulated during quasi-particle conversion in reflection and transmission processes in superconductor interfaces. This will be justified more rigorously in the following calculations:
\begin{eqnarray}
\Psi_e &=& \frac{1}{\sqrt{2}} e^{-i\phi/2} \left( \gamma_1 + i\gamma_2 \right) \;, \\
\Psi_h &=& \frac{1}{\sqrt{2}} e^{+i\phi/2} \left( \gamma_1 - i\gamma_2 \right)\;,
\end{eqnarray}
where $\Psi_e$ and $\Psi_h$ denote the electron and hole operators, respectively. The inverse transformation is given by
\begin{eqnarray}
\label{M1}
\gamma_1 &=& \frac{1}{\sqrt{2}}\left( e^{i\phi/2}\Psi_e + e^{-i\phi/2}\Psi_h \right)\;, \\
\label{M2}
i\gamma_2 &=& \frac{1}{\sqrt{2}}\left( e^{i\phi/2}\Psi_e - e^{-i\phi/2}\Psi_h \right)\;.
\end{eqnarray}

A topological insulator is characterized by the presence of a bulk energy gap. Within this gap, and at sufficiently low energies, a pair of QAH edge bands with linear dispersion emerges. In the MTI slab, two chiral QAH modes are present: one with wavenumber $k>0$ and another with $k<0$, propagating in opposite directions along opposite edges. When electron–hole symmetry is taken into account, the spectrum is effectively doubled, which must be considered when evaluating Andreev reflection processes at the normal–superconductor interface.

\subsection{NSN junction}

The scattering matrix of a NSN junction can be used as the basic building block to construct the scattering matrix of the more general NSNSN junction. It relates the amplitudes of the incoming modes to those of the outgoing modes (see Fig.~\ref{F0B}a). At sufficiently low energies, four propagating QAH modes are incident on the NSN junction: two incoming modes from the left (l) and two from the right (r), all propagating toward the superconducting region. These can be written as $a = (a_{le}, a_{lh}, a_{re}, a_{rh})^T$. Similarly, we define four outgoing modes, with two propagating toward the left contact and two toward the right contact, $b = (b_{le}, b_{lh}, b_{re}, b_{rh})^T$. The incoming and outgoing amplitudes are related by the scattering relation $b = Sa$.

We focus on the case where the superconducting regions are in the topological $\mathcal{N}=1$ phase at energies close to zero. In this regime, transport within the superconducting segments is mediated by a single chiral Majorana mode propagating along each edge. By contrast, the $\mathcal{N}=0$ and $\mathcal{N}=2$ phases lead to trivial transport behavior. In the $\mathcal{N}=0$ phase, incident QAH modes are completely reflected at the interface, whereas in the $\mathcal{N}=2$ phase they are perfectly transmitted, effectively as if superconductivity were absent.

The structure of the scattering matrix for an NSN junction is determined by the presence of a single chiral Majorana mode at each edge. Accordingly, only one Majorana mode propagates within the superconducting region along the same direction as each incoming mode. Without loss of generality, we choose the Majorana mode propagating from left to right to be $\gamma_1$. The modes $\gamma_1$ and $\gamma_2$ form an orthogonal basis of charge-neutral excitations, each given by an equal superposition of electron and hole degrees of freedom (see Eqs.~\ref{M1} and \ref{M2}).

In Fig.~\ref{F0B}a, an incoming electron from the left of amplitude $a_{le}$, is split at the first NS interface into two modes: a transmitted chiral Majorana mode $\frac{1}{\sqrt{2}} e^{-i\phi/2}a_{le}\gamma_1$ and a reflected one $\frac{i}{\sqrt{2}} e^{-i\phi/2}a_{le}\gamma_2$. These Majorana modes propagate in opposite directions along opposite edges. The transmitted Majorana crosses the superconducting region and reaches the right side without further reflections. Owing to their distinct propagation paths, the two modes acquire different phases. Transforming back to the electron–hole basis, we obtain
\begin{eqnarray}
b_{r} &=& \frac{1}{2} e^{i\beta} a_{le}
\left( \Psi_e + e^{-i\phi}\, \Psi_h  \right)\;,\\
b_{l} &=& \frac{1}{2} e^{i\alpha} a_{le}
\left( \Psi_e - e^{-i\phi}\,\Psi_h  \right)\;,
\end{eqnarray}
where $\alpha$ and $\beta$ are the phases accumulated along the corresponding propagation paths, $a_{le}$ is the incoming electron amplitude of probability and $\Psi_e$ and $\Psi_h$ are the electron hole operators. 
The relation between input and output modes can be written in matrix notation as the scattering relation $b = Sa$, where we obtain
\begin{align}
\label{eq8}
t_{ee}&=+\frac{1}{2} e^{i\beta}\; , \\
\label{eq9}
t_{he}&=+\frac{1}{2} e^{i(\beta-\phi)} \; , \\
\label{eq10}
r_{ee}&=+\frac{1}{2} e^{i\alpha} \; ,\\
\label{eq11}
r_{he}&=-\frac{1}{2} e^{i(\alpha-\phi)}\; . 
\end{align}

In Eqs.~(\ref{eq8}–\ref{eq11}), transmission and reflection processes that involve a change in the quasiparticle character (from electron to hole, or vice versa) acquire an additional phase $\pm\phi$, associated with the creation or annihilation of a Cooper pair. 
This explicitly shows that the gauge phase appearing in the electron–hole operators is directly tied to the superconducting phase.
The full scattering matrix can then be constructed by imposing particle–hole symmetry and unitarity.
It is convenient to factor out a global phase $e^{i\beta}$, thereby reducing the number of independent parameters and introducing $\xi = \alpha - \beta$. The scattering matrix then reads
\begin{align}
S(\xi,\phi)&=
\frac{1}{2}\begin{pmatrix}
1 & -e^{i\phi} & e^{i\xi} & e^{i(\xi+\phi)}\\
-e^{-i\phi} & 1 & e^{i(\xi-\phi)} & e^{i\xi}\\
e^{i\xi} & e^{i(\xi+\phi)} & 1 & -e^{i\phi}\\
e^{i(\xi-\phi)} & e^{i\xi} & -e^{-i\phi} & 1
\end{pmatrix}\; .
\end{align}

\subsection{NSNSN junction}
\label{equations}

The scattering matrix of the NSNSN system is obtained by appropriately composing the scattering matrices of the simpler NSN junction \cite{Datta,Cahay1988,Serra2009,Estarellas}. As shown in Fig.~\ref{F0B}b, the input and output modes of junction 1 are denoted by $a_1 = (a_{le1}, a_{lh1}, a_{re1}, a_{rh1})^T$ and $b_1 = (b_{le1}, b_{lh1}, b_{re1}, b_{rh1})^T$. Similarly, the input and output modes of junction 2 are defined as $a_2 = (a_{le2}, a_{lh2}, a_{re2}, a_{rh2})^T$ and $b_2 = (b_{le2}, b_{lh2}, b_{re2}, b_{rh2})^T$. The two scattering matrices, $S_1$ and $S_2$, are identical except for their respective superconducting phases $\phi_1$ and $\phi_2$. 
Further details on the composition of $S_1$ and $S_2$ into the full scattering matrix $S$ are provided in Appendix~\ref{appA}.

\begin{figure}[htbp]
\centering
  \includegraphics[width=0.5\textwidth]{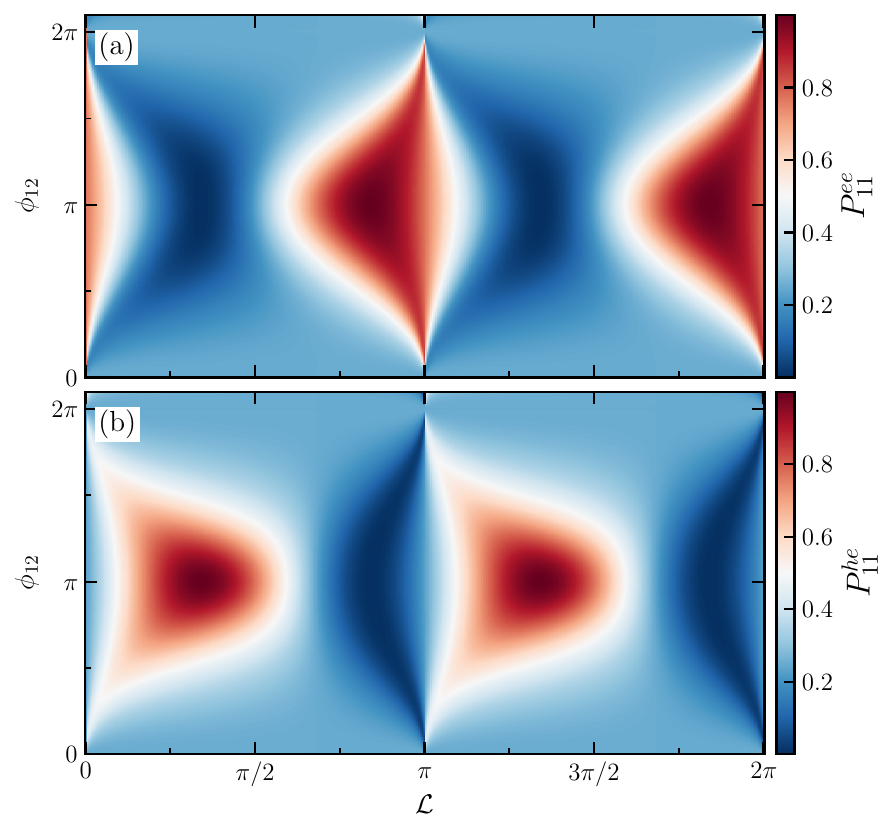}
\caption{(a) Normal reflection probability $P^{ee}_{11}$ of an NSNSN junction as a function of the phase $\mathcal{L} = k\ell$, associated with the separation between the superconducting regions, and the phase difference $\phi_{12}$. (b) Same as in (a), but for the Andreev reflection probability $P^{he}_{11}$. The parameter $\xi = 0.5$ is used.}  
  \label{F1}
\end{figure}

To calculate the local electrical and thermal conductances of the full device, we first determine the normal ($P^{ee}_{11}$) and Andreev 
($P^{he}_{11}$) reflection probabilities at the left terminal. This is done by considering a single incident electron-like QAH state incoming from the left contact, i.e., $a_{le1} = 1$ and $a_{lh1}= 0$.
We obtain the following expressions (defining $\phi_{12} = \phi_1 - \phi_2$):
\begin{widetext}
\begin{eqnarray}
P^{ee}_{11}
\equiv
|b_{le1}|^2
&=&
\frac{1}{2}
\left(
1+
\frac{
e^{-2 i \xi}\left(-1+\cos\phi_{12}\right)
}{
1-2 e^{2 i \mathcal{L}}+\cos\phi_{12})
}
\right)
\left(
\frac{1}{2}+
\frac{
e^{2 i (\xi+\mathcal{L})}\left(-1+\cos\phi_{12}\right)
}{
-4+2 e^{2 i \mathcal{L}}\left(1+\cos\phi_{12}\right)
}
\right)\; ,
\label{EQ2} \\
\rule{0cm}{1cm}
P^{he}_{11} 
\equiv
|b_{lh1}|^2
&=& 
-\frac{
\left(
e^{i(\xi+2\mathcal{L})}
-\cos\xi
- i\,\cos\phi_{12}\sin\xi
\right)
\left(
-\cos\xi
+ i\sin\xi
+ e^{2 i \mathcal{L}}
\left(
\cos\xi
- i\,\cos\phi_{12}\sin\xi
\right)
\right)
}{
\left(
1-2 e^{2 i \mathcal{L}}+\cos\phi_{12}
\right)
\left(
-2+e^{2 i \mathcal{L}}\left(1+\cos\phi_{12}\right)
\right)
}\; ,
\label{EQ3}
\end{eqnarray}

\end{widetext}
where $\mathcal{L} = k\ell$ is the phase accumulated between superconductors for modes with wavenumber $k$. 

As shown in Fig.~\ref{F1}, both $P^{ee}_{11}$ and $P^{he}_{11}$ reach their extreme values when $\phi_{12} = \pi$. These maxima and minima exhibit a periodicity of $\mathcal{L} = n\pi$. Furthermore, they occur in an alternating manner, such that a maximum in Andreev reflection corresponds to a minimum in the normal reflection probability. This behavior will be discussed in more detail in the next section. 
For $\xi = 0$, both the red (maximum) and blue (minimum) regions in the plot are symmetric with respect to $\mathcal{L} = n\pi$. However, in the more realistic case $\xi > 0$ (as shown in Fig.~\ref{F1}), this symmetry is broken and one region becomes larger than the other. In this sense, $\xi$ acts as an asymmetry parameter of the interference pattern.

\subsubsection{Mediating modes}

The two superconducting regions (1 and 2) are interconnected by two edge modes. From the analysis of an isolated NSN junction, one might expect these modes to be Majorana modes, since they are the excitations transmitted through the superconducting regions. However, the resonance induced by the coupling between the two superconducting segments leads to a more complex behavior.

We analyze these mediating modes by evaluating their charges defined as,
\begin{eqnarray}
Q_{\rightarrow} &=&  |b_{rh1}|^2 - |b_{re1}|^2\; ,\\
Q_{\leftarrow} &=&  |b_{lh2}|^2 - |b_{le2}|^2\; ,
 \end{eqnarray}
defined for the case where $a_{le1} = 1$ while $a_{lh1}=a_{le2}=a_{lh2}=0$.
We find that only the left-going mode exhibits a neutral charge, $Q_{\leftarrow} = 0$, consistent with a Majorana mode. By contrast, the charge $Q_{\rightarrow}$ of the right-going mode between the two superconducting regions (propagating along the upper edge) has a more involved expression,
\begin{widetext}
\begin{equation}
Q_{\rightarrow} =
\frac{
8\, e^{2 i (\mathcal{L}+\phi_{12})}
\sin2\mathcal{L}\,
\sin\phi_{12}
}{
\left(
e^{2 i (\mathcal{L}+\phi_1)}
+
e^{2 i (\mathcal{L}+\phi_2)}
+
2 e^{i\phi_{12}}
\left(
-2 + e^{2 i \mathcal{L}}
\right)
\right)
\left(
-4 e^{i(2\mathcal{L}+\phi_{12})}
+
\left(
e^{i\phi_1}
+
e^{i\phi_2}
\right)^2
\right)
}\;.
\label{EQ4}
\end{equation}
\end{widetext}

\begin{figure}[htbp]
  \centering
  \includegraphics[width=0.5\textwidth]{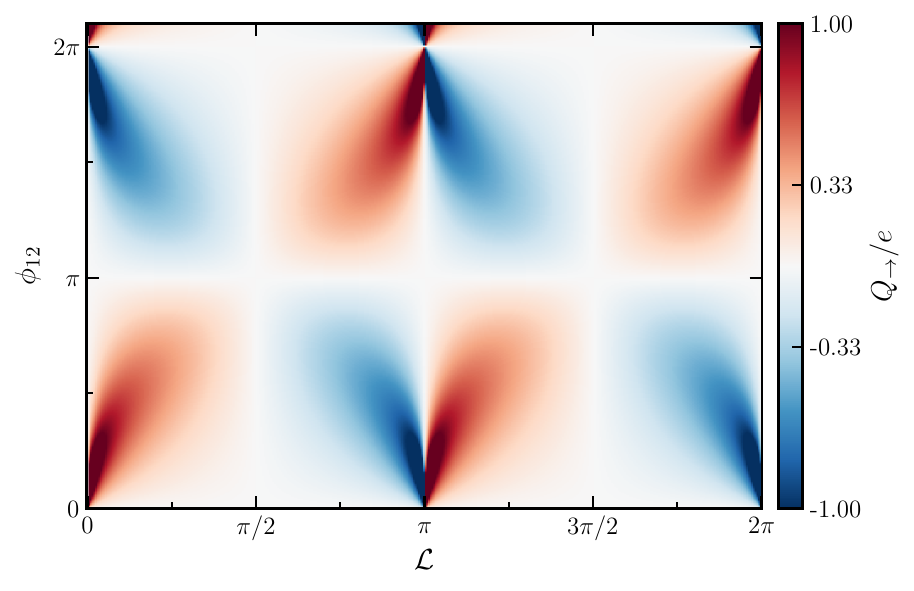}
  \caption{Charge of the right-going mode that connects the two superconductor regions as 
  a function of the phase accumulated $\mathcal{L}=k\ell$ due their separation and the phase difference $\phi_{12}\equiv\phi_1-\phi_2$ between them. Note that the charge of this mode
   is independent of $\xi$ and therefore is also independent of the superconductor length $L_s$.
  }
  \label{F3}
\end{figure}

In Fig.~\ref{F3}, we plot the character of the right-going mode $Q_{\rightarrow}$ as a function of $\phi_{12}\equiv \phi_1-\phi_2$ and $\mathcal{L}$. A periodic pattern with $\mathcal{L} = n\pi$ is observed, similar to that shown in Fig.~\ref{F1} for the normal and Andreev reflection probabilities. As $\mathcal{L}$ increases, the mode alternates between electron- and hole-like character. 
A comparison between Figs.~\ref{F1} and \ref{F3} reveals that for $\phi_{12} < \pi$, there is a correlation between the charge of the reflected mode returning to the left terminal and that of the mode mediating between the superconductors. In contrast, for $\phi_{12} > \pi$, an anti-correlation is observed.

\begin{figure}[htbp]
  \centering
  \includegraphics[width=0.5\textwidth]{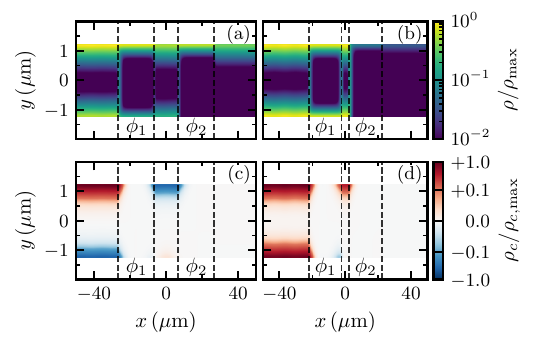}
\caption{Results from numerical modeling. Upper panels show the probability density distributions $\rho$, while the lower panels display the charge density $\rho_c$. Panels (a) and (c) correspond to $\ell = 13.3\,\mu\mathrm{m}$, whereas panels (b) and (d) correspond to $\ell = 4.4\,\mu\mathrm{m}$. 
The remaining parameters are $L_y = 2.5\,\mu\mathrm{m}$, $L_s = 20\,\mu\mathrm{m}$, $\phi_{12} = \pi/2$, $\alpha = 0.2\,\mathrm{meV}\,\mu\mathrm{m}$, $m_0 = 17\,\mathrm{meV}$, and $\hbar m_1 = 10^{-3}\,\mathrm{meV}\,\mu\mathrm{m}^2$. Additionally, $\Delta_{t} = 1\,\mathrm{meV}$ and $\Delta_{b} = 0$ in the superconducting regions.}  
  \label{F7}
\end{figure}

This correlation is further confirmed by the results shown in Fig.~\ref{F7}, where the spatial distributions of the probability and charge densities are obtained using numerical methods (see Refs.~\cite{Osca2018,Osca2019,Osca2026} and Appendix~\ref{appB}). The densities are computed for two different values of the separation length between superconducting regions $\ell$, assuming a single incident electron-like QAH mode originating from the upper-left corner of the figure.

We observe that the charge of the reflected mode at the bottom left of Figs.~\ref{F7}c and \ref{F7}d is correlated with the charge density of the mode propagating along the upper edge around $x = 0$, between the two superconducting regions. 
This charged mediating mode could, in principle, be detected using capacitance spectroscopy by placing a nearby quantum point contact. Additionally, a real-time modulation of the superconducting phase differences would induce a switching of the mode charge between positive and negative values, effectively enabling its operation as an AC charge modulator. 
By contrast, these figures also illustrate how the mode propagating along the lower edge around $x = 0$ remains essentially charge neutral, except for a small charge accumulation at the normal–superconductor interfaces.

\subsection{Onsager coefficients}

In general, the charge and heat currents $I_i$ and $J_i$ at terminal $i$ can be calculated from the reflection and transmission probabilities (see, e.g., Ref.~\cite{Daniele}). In particular, the charge ($I_1$) and heat ($J_1$) currents at terminal 1 are given by
\begin{eqnarray}
I_1 &=& \frac{e}{h} \sum_{\alpha=e,h} 
\int_0^\infty dE \,
s_\alpha \Bigl[
\,N_{1}^\alpha(E)\,f_{1}^\alpha(E) \nonumber\\
&&- \sum_{j=1,2} \sum_{\beta=e,h} P^{\alpha\beta}_{1j}(E)\,f_{j}^\beta(E)
\Bigr]\;,
\label{EQ5}\\
J_1 &=& \frac{1}{h} \sum_{\alpha=e,h} 
\int_0^\infty dE \,(E - eV_1)
\Bigl[
N_1^{\alpha}(E)\, f_1^{\alpha}(E) \nonumber\\
&&- \sum_{j=1,2} \sum_{\beta=e,h} P_{1j}^{\alpha\beta}(E)\, f_j^\beta(E)
\Bigr]\;,
\label{EQ6}
\end{eqnarray}
where $V_1$ is the bias at terminal 1, $N_1^\alpha(E)$ is the number of incoming modes of type $\alpha$ in this terminal, and $P^{\alpha\beta}_{1j}(E)$ denotes the probability that a quasiparticle of type $\beta$ in lead $j$ is transmitted as a quasiparticle of type $\alpha$ into lead 1. Furthermore, $f_i^\alpha(E)$ is the Fermi distribution at energy $E$ in terminal $i$ (with $i=1$ for the left lead and $i=2$ for the right lead) for quasiparticles of type $\alpha$. We also define $s_e = 1$ and $s_h = -1$. 

Similar equations can be written for the second terminal by swapping the subscript 1 with 2. Note that the superconductors are assumed to be grounded at a fixed bias. Therefore, charge current conservation between the two terminals is not imposed.

We assume a finite bias $V_1$ applied only to terminal 1, while terminal 2 is kept at zero bias and 
temperature $T_0$ close to zero, such that the thermal energy satisfies $k_B T_0  \ll E_g$, where $E_g$ is the band gap energy. 
The temperature difference between the terminals is defined as $\Delta T_1 = T_1 - T_0$. Linearizing Eqs.~(\ref{EQ5}) and (\ref{EQ6}), we obtain the Onsager relations \cite{Taddei,Liliana,Arra2025},
\begin{align}
I_1 &= L_{11}\,\frac{V_1}{T_0} + L_{12}\,\frac{\Delta T_1}{T_0^2}\,,
\label{EQ7a}\\
J_1 &= L_{21}\,\frac{V_1}{T_0} + L_{22}\,\frac{\Delta T_1}{T_0^2}\,.
\label{EQ7}
\end{align}

In Eqs.~(\ref{EQ7a},\ref{EQ7}), $L_{11}$, $L_{12}$, $L_{21}$, and $L_{22}$ are the 
Onsager coefficients.  They  can be expressed as
\begin{widetext}
\begin{align}
L_{11} &= \frac{e^2 T_0}{h} \int_{0}^{\infty} dE\,
\Bigl[
N_{11}^{e}(E) - P_{11}^{ee}(E) + P_{11}^{he}(E)
+ N_{11}^{h}(E) - P_{11}^{hh}(E) + P_{11}^{eh}(E)
\Bigr]
\left(-\frac{\partial f}{\partial E}\right)\;, \\[6pt]
L_{22} &= \frac{T_0}{h} \int_{0}^{\infty} dE\,E^{2}\,
\Bigl[
N_{11}^{e}(E) - P_{11}^{ee}(E) - P_{11}^{he}(E)
+ N_{11}^{h}(E) - P_{11}^{hh}(E) - P_{11}^{eh}(E)
\Bigr]
\left(-\frac{\partial f}{\partial E}\right)\;,\\[6pt]
L_{12} &= \frac{e T_0}{h^{2}} \int_{0}^{\infty} dE\,E\,
\Bigl[
N_{11}^{e}(E) - P_{11}^{ee}(E) + P_{11}^{he}(E)
- N_{11}^{h}(E) + P_{11}^{hh}(E) - P_{11}^{eh}(E)
\Bigr]
\left(-\frac{\partial f}{\partial E}\right)\;, \\[6pt]
L_{21} &= \frac{e T_0}{h^{2}} \int_{0}^{\infty} dE\,E\,
\Bigl[
N_{11}^{e}(E) - P_{11}^{ee}(E) - P_{11}^{he}(E)
- N_{11}^{h}(E) + P_{11}^{hh}(E) + P_{11}^{eh}(E)
\Bigr]
\left(-\frac{\partial f}{\partial E}\right)\;.
\label{EQ8}
\end{align}
\end{widetext}
In a superconductor-mediated device at zero or near-zero energies, electron–hole symmetry enforces $L_{12} = L_{21} = 0$.  In this case, $L_{11}$ and $L_{22}$ simplify to 
\begin{eqnarray}
L_{11} &=& \frac{e^2 T_0}{h}
\displaystyle\int_{-\infty}^{\infty} dE\,
\Bigl[
N_{11}^{e}(E) \nonumber\\
&-& P_{11}^{ee}(E) + P_{11}^{he}(E)
\Bigr]
\left(-\frac{\partial f}{\partial E}\right), \\
L_{22} &=& \frac{T_0}{h} \int_{-\infty}^{\infty} dE\,E^{2}
\Bigl[
N_{11}^{e}(E)\nonumber\\
&-& P_{11}^{ee}(E) - P_{11}^{he}(E)
\Bigr]
\left(-\frac{\partial f}{\partial E}\right).
\label{EQ9}
\end{eqnarray}

At low temperatures ($K_B T_0 \ll E_g$), a Sommerfeld expansion \cite{Ashcroft76} yields
\begin{align}
L_{11} &= \frac{e^2 T_0}{h} \left[ N_{11}^{e}(0) - P_{11}^{ee}(0) + P_{11}^{he}(0) \right], \\[6pt]
L_{22} &= \frac{T_0}{h}\,\frac{\pi^2}{3} (k_B T_0)^2 \left[ N_{11}^{e}(0) - P_{11}^{ee}(0) - P_{11}^{he}(0) \right].
\label{EQ10}
\end{align}
As a consequence, the electrical conductance $G_1=I_1/V_1$ and the thermal conductance $K_1=J_1/\Delta{T_1}$ at terminal 1 are written as,
\begin{eqnarray}
\label{EQ11}
G_1 &=& \frac{e^2 }{h} \left( N_{11}^{e}(0) - P_{11}^{ee}(0) + P_{11}^{he}(0)\right)\;, \\
\label{EQ12}
K_1 &=& \frac{\pi^2}{3} \frac{k_B^2}{h}T_0  \left( N_{11}^{e}(0) - P_{11}^{ee}(0) - P_{11}^{he}(0)\right)\; . 
\end{eqnarray}

Note that no Seebeck or Peltier effects are present in this device at first order, since $L_{12} = L_{21} = 0$. Substituting Eqs.~(\ref{EQ2}) and (\ref{EQ3}) into Eqs.~(\ref{EQ11}) and (\ref{EQ12}), we obtain
\begin{widetext}
\begin{eqnarray}
\label{EQ13}
\frac{G_1}{G_0} &=& \frac{1}{2} e^{-2 i \xi} \Bigg[ 
- \big(-1 + e^{2 i \xi}\big)^2
- \frac{2 \big(-1 + e^{2 i \mathcal{L}}\big)}{1 - 2 e^{2 i \mathcal{L}} + \cos\phi_{12}}
+ \frac{2 e^{4 i \xi} \big(-1 + e^{2 i \mathcal{L}}\big)}{-2 + e^{2 i \mathcal{L}} \big(1 + \cos\phi_{12}\big)} 
\Bigg]\;,\\
\label{EQ14}
\frac{K_1}{K_0} &=& -\frac{ \left(-1 + e^{2 i \mathcal{L}}\right)^2 \left[1 + \cos\phi_{12}\right] }{ \left[1 - 2 e^{2 i \mathcal{L}} + \cos\phi_{12}\right] \left[-2 + e^{2 i \mathcal{L}} \left(1 + \cos\phi_{12}\right)\right] }\; ,
\end{eqnarray}
\end{widetext}
where $G_0 = e^2/h$ and $K_0 = {\pi^2 k_B^2 T_0}/{3h}$. All other parameters are defined as in the previous expressions.

\begin{figure}[htbp]
  \centering
  \includegraphics[width=0.5\textwidth]{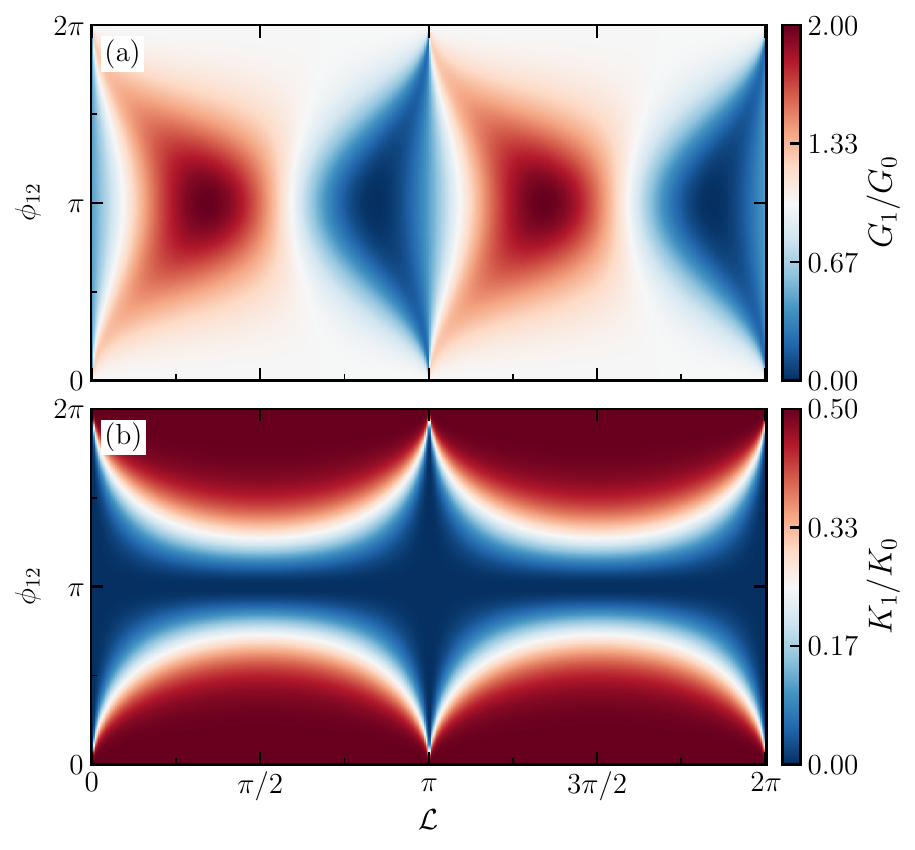}
\caption{Electrical (a) and thermal (b) conductances of the NSNSN junction in the linear regime, shown as functions of the propagation phase $\mathcal{L}$ and the phase difference $\phi_{12} \equiv \phi_1 - \phi_2$. The parameter $\xi = 0.5$ is used.}
\label{F2}
\end{figure}

Figure~\ref{F2} shows the electrical conductance $G$ and the thermal conductance $K$ as functions of the phase difference $\phi_{12}$ and the propagation phase $\mathcal{L}$. The proportionality of the thermal conductance (see Eq.~\ref{EQ12}) to the quasiparticle current at near-zero energy provides a direct indication of the quasiparticle flow through the device. 

As shown in Fig.~\ref{F2}b, the system exhibits quasiparticle current blockades with $\pi$ periodicity in both $\phi_{12}$ and $\mathcal{L}$. In these regimes, the right superconducting region becomes effectively opaque to modes incoming from the left, thereby disconnecting the two terminals. As a result, two Majorana modes are reflected back toward the left contact: one arising from scattering in the left superconducting region and the other in the right region. These reflected Majorana modes interfere, with the resulting interference pattern depending on the difference in their propagation paths. This leads to a reflected electron- or hole-like excitation (or a superposition of both), as shown in Fig.~\ref{F2}a.

The conductance alternates between maxima ($G_1/G_0 = 2$) and minima ($G_1/G_0 = 0$) as $\mathcal{L}$ increases, particularly for $\phi_{12} = \pi$  it is $G_1/G_0=2\sin(\xi+\mathcal{L})$ and $K_1/K_0=0$. By comparing Fig.~\ref{F2} with Fig.~\ref{F1}, it is evident that the conductance maxima correspond to full Andreev reflection driven by Majorana interference, whereas the minima correspond to full normal reflection.
On the other hand  if $\phi_{12} = 0, 2\pi$ the result is ideal single Majorana transmission, $G_1/G_0=1$ and $K_1/K_0=0.5$.

Although the alternating pattern is centered around $\mathcal{L} = n\pi$, it is not perfectly symmetric. The shapes of the minima (blue regions) and maxima (red regions) differ, and this asymmetry is controlled by the parameter $\xi$, which is related to the length of the superconducting segments. The pattern becomes symmetric when $\xi = n\pi$.

As a side result, we find that the Andreev reflection probability can be directly related to the conductances $G_1$ and $K_1$. This relation can be verified by comparison with the results shown in Fig.~\ref{F1}b,
\begin{equation}
P_{11}^{he}(0) =
\frac{1}{2}\left( \frac{G_1}{G_0} - \frac{K_1}{K_0} \right).
\end{equation}

\begin{figure}[htbp]
  \centering
  \includegraphics[width=0.5\textwidth]{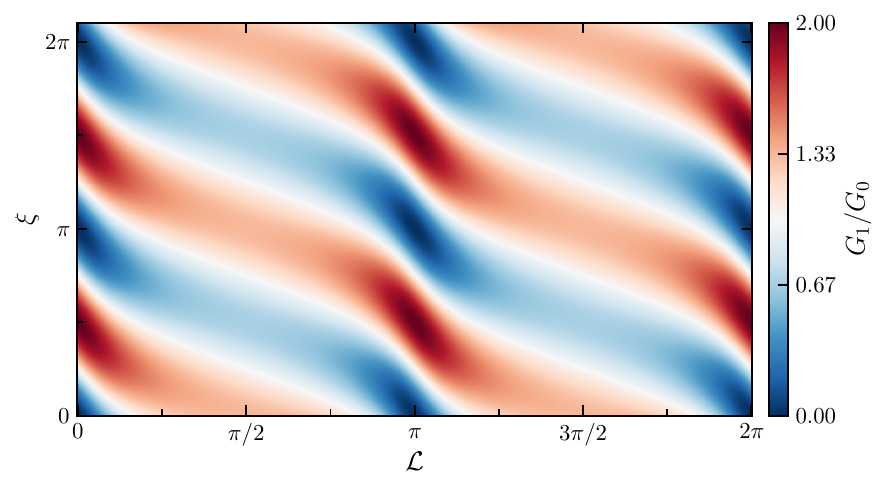}
\caption{Electrical conductance of the NSNSN junction in the linear regime as a function of the propagation phase $\mathcal{L} = k\ell$, associated with the separation between superconducting regions, and the parameter $\xi$, which is proportional to the superconducting segment length $L_s$. The phase difference is fixed to $\phi_{12} = \pi$.}
\label{F4}
\end{figure}

Figure~\ref{F4} shows the conductance $G_1$ as a function of the phase $\mathcal{L}$ and the parameter $\xi$. Since $\xi \propto L_s$, this plot illustrates the role of the superconducting segment length $L_s$ in determining the magnitude of the electrical conductance. We assume that both superconducting regions have equal length.
In this case, $\pi$-periodic quasiparticle current blockades occur only when $\mathcal{L} = n\pi$ (see Fig.~\ref{F4}), leading to an alternation between maxima and minima of $G_1$ as $\xi$ increases. The propagation paths of the reflected Majorana modes are then set by the lengths of the superconducting regions.

In general, the reflected state may have predominantly electron- or hole-like character depending on $\mathcal{L}$. Limiting  values $G_1/G_0 = 2$ or $G_1/G_0 = 0$ are reached at values $\mathcal{L} = n\pi$. The alternation between electron- and hole-like regimes appears as diagonal stripe patterns in Fig.~\ref{F4}. These results are consistently reproduced by the numerical model, as discussed in Appendix~\ref{appB}.

In general, we are interpreting $\xi$ and $\mathcal{L}$ as values proportional to $L_s$ and $\ell$ respectively. While this is true, both quantities are also proportional to $k$ which is also proportional to energy because
of the linear dispersion relation of the topological modes. Therefore, increasing $\xi$ and $\mathcal{L}$ can be also associated to increments in energy provided we do not deviate too much from zero. As the energy increases the topological mode deviates further from an ideal Majorana state, rendering the underlying approximations of the analytical model progressively less accurate.

\subsection{The AC Majorana effect}

The phase-controlled modulation of the conductance $G_1$ provides a distinctive electrical signature of the ellusive chiral Majorana quasiparticles. We refer to this phenomenon as the AC Majorana effect.
According to the second Josephson relation, in the presence of a voltage bias $V_b$ between two superconductors, their phase difference evolves in time as
\begin{equation}
\phi_{12}(t) = \phi_0 + \frac{2eV_b}{\hbar} t.
\label{AC1}
\end{equation}
As a result, the phase-dependent conductance $G_1(\phi_{12})$ is converted into a time-dependent signal $G_1(t)$.
This mechanism suggests a feasible route for experimental detection via time-dependent conductance measurements. In practice, the modulation can be achieved by applying a small voltage bias $V_b$ to one superconducting island while keeping the other grounded, thereby inducing a dynamical phase difference.
In the adiabatic regime (defined by $2eV_b \ll E_g$, with $E_g$ the energy gap) the conductance oscillates as given by Eq.~(\ref{EQ13}) with $\phi_{12}$ from Eq.~(\ref{AC1}). This oscillatory behavior constitutes a direct and experimentally accessible signature of Majorana interference in topological superconducting devices.

The AC Majorana effect shares a formal similarity with the conventional AC Josephson effect, since both arise from the time evolution of the superconducting phase difference according to the Josephson relation. In both cases, a constant voltage bias induces an oscillatory response with frequency $2eV/\hbar$. 
However, the physical origin and observable signatures of the two effects are fundamentally different. The AC Josephson effect manifests as an oscillating supercurrent carried by Cooper pairs, whereas the AC Majorana effect corresponds to a modulation of the conductance arising from the interference of chiral Majorana quasiparticles. Unlike the Josephson effect, which reflects condensate coherence, the Majorana AC response is inherently an interferometric phenomenon involving quasiparticle  transport along topological edge states.
The conductance oscillations predicted here can exhibit strong non-sinusoidal behavior, including complete suppression or enhancement,
depending on the phase ${\cal L}$  in Fig.\ \ref{F2}.
The detailed $G_1(t)$ AC shapes, contained in the analytical result
Eq.\ (\ref{EQ13}), are additional signatures of this peculiar interference mechanism.

\section{Three terminals}
\label{sec3}

\begin{figure}[htbp]
  \centering
  \includegraphics[width=0.4\textwidth]{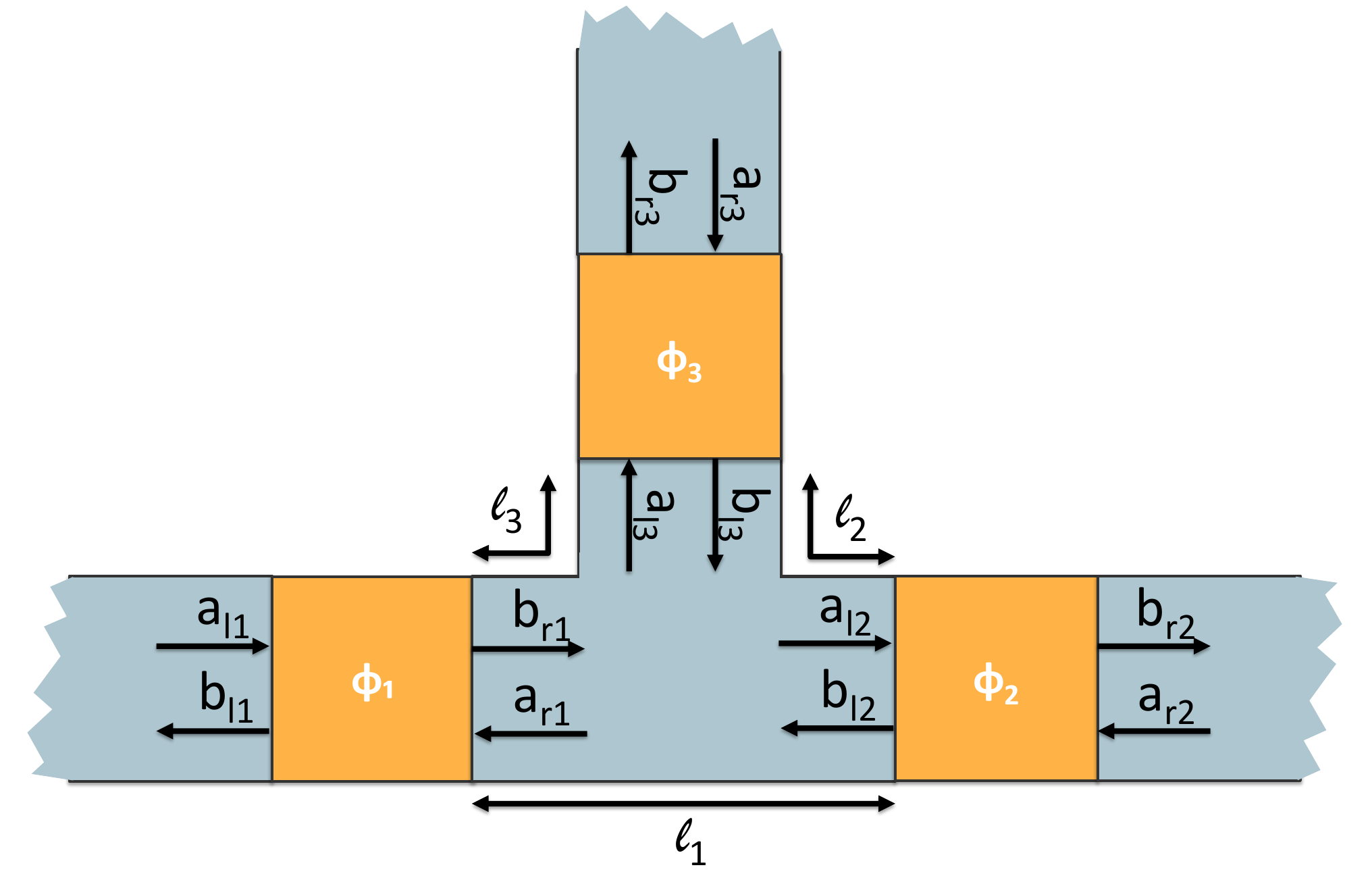}
  \caption{Diagram of a triple junction.}
  \label{F10}
\end{figure}

Here, we consider a three-terminal device with normal contacts, where each arm contains a region that is proximity coupled to a superconductor with a distinct phase $\phi$. The central connecting region, where the three arms meet, is not proximity coupled and therefore remains in the normal state. 
A schematic of the system, indicating the superconducting phases of the three islands, the corresponding input and output amplitudes, and the distances between them, is shown in Fig.~\ref{F10}.

The composition of the scattering matrices must now account for the distinct interconnections generated by the edge modes between the different contacts. Due to the chirality of the device, outgoing modes propagate clockwise: modes leaving terminal 3 are directed toward terminal 2, those from terminal 2 propagate to terminal 1, and modes from terminal 1 return to terminal 3. 

In this section, we consider terminal 3 to be biased at voltage $V_3$ and temperature $T_3 > T_0$, while terminals 1 and 2 are kept at zero bias and temperature $T_0$ (close to zero, $k_B T_0\ll E_g$). We solve the corresponding system of equations assuming that the only incident modes are electron-like QAH modes injected from the upper terminal.

With this configuration, the system effectively involves two reservoirs: one at finite bias and elevated temperature, and another at zero bias and low temperature $T_0$. Accordingly, the expressions for the electrical and thermal conductances, $G_3$ and $K_3$, at terminal 3 can be written in analogy with Eqs.~(\ref{EQ11}) and (\ref{EQ12}) by replacing the indices $1 \rightarrow 3$. Similarly, Eqs.~(\ref{EQ5}) and (\ref{EQ6}) also hold for the new terminals $1$ and $2$ with the corresponding index swap.

The resulting system of equations yields a new set of outgoing mode amplitudes and, consequently, a new set of scattering probabilities. For the three-terminal device,
\begin{eqnarray}
\frac{G_3}{G_0} = 
1  &+& \frac{1}{4} \left( \mathcal{I}_+ \mathcal{I}_+^*
- \mathcal{I}_- \mathcal{I}_-^* \right) \; , \\
\frac{K_3}{K_0} = 1 &-& \frac{1}{4} \left( \mathcal{I}_+ \mathcal{I}_+^* +  \mathcal{I}_- \mathcal{I}_-^* \right) .
\end{eqnarray}

where 

\begin{eqnarray}
\mathcal{I}_\pm&=&\left( 1\pm
\frac{
8 \cos\tilde\phi_{12}\,
  \sin\tilde\phi_{13}\,
  \sin\tilde\phi_{23}\,
  e^{i\Phi}
}{
\mathcal{D}
}
\right)\; ,\\
\tilde\phi_{ij}&=&\frac{\phi_i-\phi_j}{2}\; ,\\
\rule{0cm}{0.5cm}
\Phi &=& 
2\xi+\mathcal{L}_T
+\phi_1+\phi_2+\phi_3\; , \\
\rule{0cm}{0.5cm}
\mathcal{D} &=&
e^{i(\mathcal{L}_T+2\phi_1+\phi_2)}
+
e^{i(\mathcal{L}_T+\phi_1+2\phi_2)}
+
e^{i(\mathcal{L}_T+2\phi_1+\phi_3)}
\nonumber\\
&+&
e^{i(\mathcal{L}_T+2\phi_2+\phi_3)}
+
e^{i(\mathcal{L}_T+\phi_1+2\phi_3)}
+
e^{i(\mathcal{L}_T+\phi_2+2\phi_3)}
\nonumber\\
\rule{0cm}{0.5cm}
&-&
8 e^{i(\phi_1+\phi_2+\phi_3)}
+
2 e^{i(\mathcal{L}_T+\phi_1+\phi_2+\phi_3)}\; .
\end{eqnarray}
In these expressions, $\mathcal{L}_T = k(\ell_1 + \ell_2 + \ell_3)$. As shown in Fig.~\ref{F10}, $\ell_1$ denotes the distance between superconducting regions (1) and (2), while $\ell_2$ is the distance traveled by an edge mode propagating between regions (3) and (2), and $\ell_3$ corresponds to the distance between superconducting regions (1) and (3). 
Thus, $\ell_1 + \ell_2 + \ell_3$ represents the total path length of a chiral mode that originates from superconducting island 3 and returns to it.

\begin{figure}[t]
  \centering
  \includegraphics[width=0.5\textwidth]{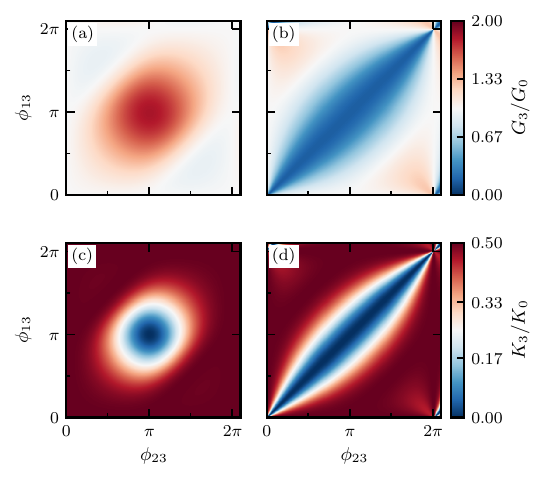}
  \caption{a) Electrical conductance on the linear regime  of the triple junction as a function of the phase difference between superconductors $\phi_{13}\equiv\phi_1-\phi_3$ and $\phi_{23}\equiv\phi_2-\phi_3$. $\mathcal{L}_1=\pi/2$ is used.  b) The same as a but with  $\mathcal{L}_1=\pi$. c) Thermal conductance in the linear regime as a function of the phase difference between superconductors $\phi_{13}$ and $\phi_{23}$.  $\mathcal{L}_1=\pi/2$ is used in this case. d) The same as c but with  $\mathcal{L}_1=\pi$. In all plots $\mathcal{L}_2=\mathcal{L}_3=\mathcal{L}_1/2$.
   }
  \label{F8}
\end{figure}

\begin{figure}[t]
  \centering
  \includegraphics[width=0.3\textwidth]{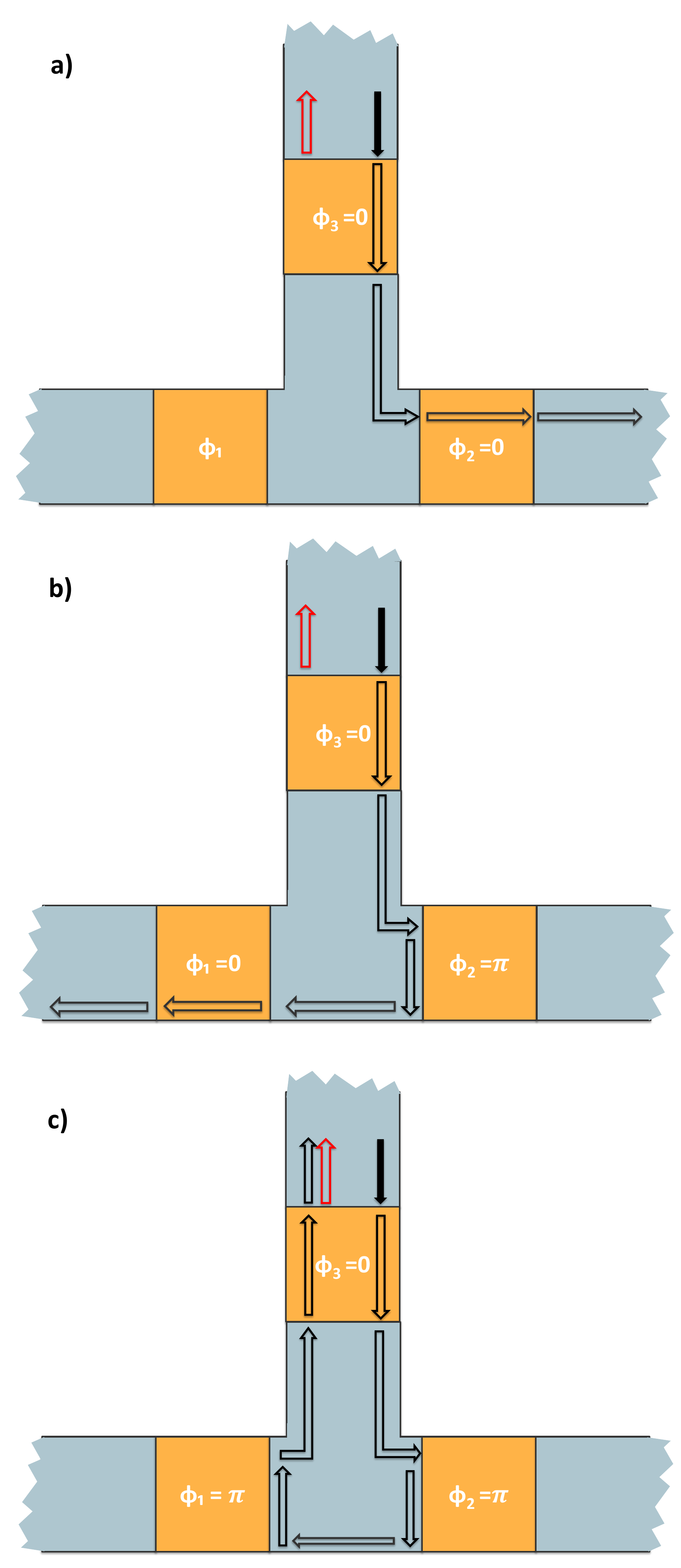}
\caption{Schematic of the transport processes in a three-terminal device. Filled black arrows represent incident electron-like QAH modes, while open arrows denote chiral Majorana edge states.}
\label{F9}
\end{figure}

Figure~\ref{F8} shows the dependence of both the electrical and thermal conductances at terminal 3 on the superconducting phase differences $\phi_{13}$ and $\phi_{23}$, while the remaining parameters are kept fixed. The results are presented for two different values of the propagation phase $\mathcal{L}_1$, assuming $\mathcal{L}_2 = \mathcal{L}_3 = \mathcal{L}_1/2$. 
The thermal conductance in the linear regime is proportional to the quasiparticle current, and therefore provides a direct measure of the transport through the device. A minimum in the thermal conductance (and thus in the quasiparticle current) occurs for $\phi_{13} = \phi_{23} = \pi$. In this situation, an incident electron-like QAH state from terminal 3 splits into two Majorana modes: one is reflected back to terminal 3, while the other is transmitted toward terminal 2. 
As illustrated in the schematic of Fig.~\ref{F9}c, this transmitted Majorana is subsequently reflected at terminal 2 and then again at terminal 1, eventually returning to the upper terminal. This behavior arises because both superconducting islands are phase-shifted by $\pi$ with respect to the superconductor connected to terminal 3. 
In this regime, the three-terminal device exhibits a complete quasiparticle current blockade, analogous to the two-terminal case for $\phi_1 - \phi_2 = \pi$. Consistently, one finds that the thermal conductances 
vanish at all three terminals $K_1 = K_2 = K_3 = 0$ for these particular values of the superconducting phases.

The above blockade is lifted as soon as one of the superconducting phase differences deviates from $\pi$. Interestingly, the way in which the blockade is lifted depends sensitively on the value of $\mathcal{L}_1$.
For the particular case $\mathcal{L}_1 = \pi/2$, the blockade occurs around the point $\phi_{13} = \phi_{23} = \pi$ (see Fig.~\ref{F8}). 
In this regime, the interference between the two reflected Majorana modes gives rise to a hole-like QAH excitation that propagates back into terminal 3, resulting in a maximum of the conductance at this terminal.
By contrast, for $\mathcal{L}_1 = \pi$, the blockade extends to the more general condition $\phi_{13} = \phi_{23}$. In this regime, the interference between the two Majorana modes reflected toward the upper terminal produces an electron-like QAH excitation. Consequently, this leads to a minimum of the conductance at terminal 3.

In general, this device operates as a Majorana router. When $\phi_{23} = 0$ and $\phi_{13}$ is arbitrary, one Majorana mode is reflected back to terminal 3, while the other is routed to terminal 2 (see Fig.~\ref{F9}a). In this configuration, the thermal conductances take the values $K_3 = K_2 = 0.5\,K_0$, while $K_1 = 0$.
Conversely, when $\phi_{23} = \pi$ and $\phi_{13} = 0$ (as shown in Fig.~\ref{F9}b), one Majorana is again reflected back to terminal 3, while the second is transmitted to terminal 1. In this case, $K_3 = K_1 = 0.5\,K_0$, with $K_2 = 0$. Finally, when both phase differences satisfy $\phi_{13}= \phi_{23} = \pi$, the system enters the blockade regime as presented in Fig.~\ref{F9}c and discussed above.

\section{Conclusions}
\label{sec4}

We have obtained the scattering matrix of an NSN topological junction
hosting a chiral Majorana mode
and used it to construct the corresponding scattering matrices of both (NSNSN) two-terminal and three-terminal junctions. The analytical model has been validated against an equivalent numerical approach in the regime where tunneling is negligible, and we have shown that the predicted results remain robust for small deviations of the energy from zero.
An analysis of normal and Andreev reflection processes reveals that the primary role of the first encountered superconducting region is to split an incident electron-like mode into two chiral Majorana modes: one reflected and one transmitted. The transmitted Majorana mode may, in turn, be either reflected or transmitted at subsequent superconducting regions, depending on the relative superconducting phases. 

A fully reflecting regime can be achieved in both two- and three-terminal configurations. In this regime, two reflected Majorana modes interfere, giving rise to a QAH excitation with either electron-like or hole-like character, which directly affects the conductance. The nature of the reflected mode is determined by the difference in the propagation paths of the interfering Majoranas. To the best of our knowledge, this type of interference is a unique signature of Majorana modes.
A particularly notable outcome of our analysis is the prediction of an AC Majorana effect arising from the time dependence of superconducting phase differences. This dynamical response thus offers a promising route for the detection and manipulation of chiral Majorana modes in proximitized magnetic topological insulators.

The three-terminal geometry introduces additional richness to this blockade phenomenon, with additional phase configurations that produce a full quasiparticle current suppression, depending on the device geometry. We have demonstrated that the proposed three-terminal setup can operate as a phase-controlled Majorana router, enabling the controlled routing of Majorana excitations to a selected terminal through appropriate tuning of the superconducting phase differences.
\vfill

\begin{acknowledgments}

This project is financially supported
by MCIU/AEI/10.13039/501100011033 under Project No.
PCI2022-132927 of the QuantERA grant MAGMA, Project
No. PID2023-151975NB-I00, project No. 
CEX2021-001164-M (Maria de Maeztu), and by the European Union
NextGenerationEU/PRTR and FEDER.

\end{acknowledgments} 

\bibliography{Article4bib}

\begin{thebibliography}{36}%
\makeatletter
\providecommand \@ifxundefined [1]{%
 \@ifx{#1\undefined}
}%
\providecommand \@ifnum [1]{%
 \ifnum #1\expandafter \@firstoftwo
 \else \expandafter \@secondoftwo
 \fi
}%
\providecommand \@ifx [1]{%
 \ifx #1\expandafter \@firstoftwo
 \else \expandafter \@secondoftwo
 \fi
}%
\providecommand \natexlab [1]{#1}%
\providecommand \enquote  [1]{``#1''}%
\providecommand \bibnamefont  [1]{#1}%
\providecommand \bibfnamefont [1]{#1}%
\providecommand \citenamefont [1]{#1}%
\providecommand \href@noop [0]{\@secondoftwo}%
\providecommand \href [0]{\begingroup \@sanitize@url \@href}%
\providecommand \@href[1]{\@@startlink{#1}\@@href}%
\providecommand \@@href[1]{\endgroup#1\@@endlink}%
\providecommand \@sanitize@url [0]{\catcode `\\12\catcode `\$12\catcode
  `\&12\catcode `\#12\catcode `\^12\catcode `\_12\catcode `\%12\relax}%
\providecommand \@@startlink[1]{}%
\providecommand \@@endlink[0]{}%
\providecommand \url  [0]{\begingroup\@sanitize@url \@url }%
\providecommand \@url [1]{\endgroup\@href {#1}{\urlprefix }}%
\providecommand \urlprefix  [0]{URL }%
\providecommand \Eprint [0]{\href }%
\providecommand \doibase [0]{https://doi.org/}%
\providecommand \selectlanguage [0]{\@gobble}%
\providecommand \bibinfo  [0]{\@secondoftwo}%
\providecommand \bibfield  [0]{\@secondoftwo}%
\providecommand \translation [1]{[#1]}%
\providecommand \BibitemOpen [0]{}%
\providecommand \bibitemStop [0]{}%
\providecommand \bibitemNoStop [0]{.\EOS\space}%
\providecommand \EOS [0]{\spacefactor3000\relax}%
\providecommand \BibitemShut  [1]{\csname bibitem#1\endcsname}%
\let\auto@bib@innerbib\@empty
\bibitem [{\citenamefont {Fu}\ and\ \citenamefont {Kane}(2008)}]{Fu2008}%
  \BibitemOpen
  \bibfield  {author} {\bibinfo {author} {\bibfnamefont {L.}~\bibnamefont
  {Fu}}\ and\ \bibinfo {author} {\bibfnamefont {C.~L.}\ \bibnamefont {Kane}},\
  }\bibfield  {title} {\bibinfo {title} {Superconducting proximity effect and
  {{M}ajorana} fermions at the surface of a topological insulator},\ }\href
  {https://doi.org/10.1103/PhysRevLett.100.096407} {\bibfield  {journal}
  {\bibinfo  {journal} {Phys. Rev. Lett.}\ }\textbf {\bibinfo {volume} {100}},\
  \bibinfo {pages} {096407} (\bibinfo {year} {2008})}\BibitemShut {NoStop}%
\bibitem [{\citenamefont {Qi}\ and\ \citenamefont {Zhang}(2011)}]{Qi2011}%
  \BibitemOpen
  \bibfield  {author} {\bibinfo {author} {\bibfnamefont {X.-L.}\ \bibnamefont
  {Qi}}\ and\ \bibinfo {author} {\bibfnamefont {S.-C.}\ \bibnamefont {Zhang}},\
  }\bibfield  {title} {\bibinfo {title} {Topological insulators and
  superconductors},\ }\href {https://doi.org/10.1103/RevModPhys.83.1057}
  {\bibfield  {journal} {\bibinfo  {journal} {Rev. Mod. Phys.}\ }\textbf
  {\bibinfo {volume} {83}},\ \bibinfo {pages} {1057} (\bibinfo {year}
  {2011})}\BibitemShut {NoStop}%
\bibitem [{\citenamefont {Sato}\ and\ \citenamefont
  {Fujimoto}(2016)}]{Sato2016}%
  \BibitemOpen
  \bibfield  {author} {\bibinfo {author} {\bibfnamefont {M.}~\bibnamefont
  {Sato}}\ and\ \bibinfo {author} {\bibfnamefont {S.}~\bibnamefont
  {Fujimoto}},\ }\bibfield  {title} {\bibinfo {title} {{M}ajorana fermions and
  topology in superconductors},\ }\href@noop {} {\bibfield  {journal} {\bibinfo
   {journal} {J. Phys. Soc. Jpn.}\ }\textbf {\bibinfo {volume} {85}},\ \bibinfo
  {pages} {072001} (\bibinfo {year} {2016})}\BibitemShut {NoStop}%
\bibitem [{\citenamefont {Sato}\ and\ \citenamefont {Ando}(2017)}]{Sato2017}%
  \BibitemOpen
  \bibfield  {author} {\bibinfo {author} {\bibfnamefont {M.}~\bibnamefont
  {Sato}}\ and\ \bibinfo {author} {\bibfnamefont {Y.}~\bibnamefont {Ando}},\
  }\bibfield  {title} {\bibinfo {title} {Topological superconductors: a
  review},\ }\href {https://doi.org/10.1088/1361-6633/aa6ac7} {\bibfield
  {journal} {\bibinfo  {journal} {Reports on Progress in Physics}\ }\textbf
  {\bibinfo {volume} {80}},\ \bibinfo {pages} {076501} (\bibinfo {year}
  {2017})}\BibitemShut {NoStop}%
\bibitem [{\citenamefont {He}\ \emph {et~al.}(2019)\citenamefont {He},
  \citenamefont {Liang}, \citenamefont {Tanaka},\ and\ \citenamefont
  {Nagaosa}}]{He2019}%
  \BibitemOpen
  \bibfield  {author} {\bibinfo {author} {\bibfnamefont {J.~J.}\ \bibnamefont
  {He}}, \bibinfo {author} {\bibfnamefont {T.}~\bibnamefont {Liang}}, \bibinfo
  {author} {\bibfnamefont {Y.}~\bibnamefont {Tanaka}},\ and\ \bibinfo {author}
  {\bibfnamefont {N.}~\bibnamefont {Nagaosa}},\ }\bibfield  {title} {\bibinfo
  {title} {Platform of chiral majorana edge modes and its quantum transport
  phenomena},\ }\href {https://doi.org/10.1038/s42005-019-0250-5} {\bibfield
  {journal} {\bibinfo  {journal} {Communications Physics}\ }\textbf {\bibinfo
  {volume} {2}},\ \bibinfo {pages} {149} (\bibinfo {year} {2019})}\BibitemShut
  {NoStop}%
\bibitem [{\citenamefont {Rachel}\ \emph {et~al.}(2017)\citenamefont {Rachel},
  \citenamefont {Mascot}, \citenamefont {Cocklin}, \citenamefont {Vojta},\ and\
  \citenamefont {Morr}}]{Rachel}%
  \BibitemOpen
  \bibfield  {author} {\bibinfo {author} {\bibfnamefont {S.}~\bibnamefont
  {Rachel}}, \bibinfo {author} {\bibfnamefont {E.}~\bibnamefont {Mascot}},
  \bibinfo {author} {\bibfnamefont {S.}~\bibnamefont {Cocklin}}, \bibinfo
  {author} {\bibfnamefont {M.}~\bibnamefont {Vojta}},\ and\ \bibinfo {author}
  {\bibfnamefont {D.~K.}\ \bibnamefont {Morr}},\ }\bibfield  {title} {\bibinfo
  {title} {Quantized charge transport in chiral majorana edge modes},\ }\href
  {https://doi.org/10.1103/PhysRevB.96.205131} {\bibfield  {journal} {\bibinfo
  {journal} {Phys. Rev. B}\ }\textbf {\bibinfo {volume} {96}},\ \bibinfo
  {pages} {205131} (\bibinfo {year} {2017})}\BibitemShut {NoStop}%
\bibitem [{\citenamefont {Lao}\ and\ \citenamefont {Zhou}(2024)}]{Lao}%
  \BibitemOpen
  \bibfield  {author} {\bibinfo {author} {\bibfnamefont {J.}~\bibnamefont
  {Lao}}\ and\ \bibinfo {author} {\bibfnamefont {T.}~\bibnamefont {Zhou}},\
  }\bibfield  {title} {\bibinfo {title} {Manipulating chiral majorana mode with
  additional potential in superconductor-chern insulator heterostructures},\
  }\href {https://doi.org/10.1088/1361-648X/ad5e2c} {\bibfield  {journal}
  {\bibinfo  {journal} {Journal of Physics: Condensed Matter}\ }\textbf
  {\bibinfo {volume} {36}},\ \bibinfo {pages} {405702} (\bibinfo {year}
  {2024})}\BibitemShut {NoStop}%
\bibitem [{\citenamefont {Liang}\ \emph {et~al.}(2012)\citenamefont {Liang},
  \citenamefont {Wang},\ and\ \citenamefont {Hu}}]{Liang}%
  \BibitemOpen
  \bibfield  {author} {\bibinfo {author} {\bibfnamefont {Q.-F.}\ \bibnamefont
  {Liang}}, \bibinfo {author} {\bibfnamefont {Z.}~\bibnamefont {Wang}},\ and\
  \bibinfo {author} {\bibfnamefont {X.}~\bibnamefont {Hu}},\ }\bibfield
  {title} {\bibinfo {title} {Manipulation of majorana fermions by point-like
  gate voltage in the vortex state of a topological superconductor},\ }\href
  {https://doi.org/10.1209/0295-5075/99/50004} {\bibfield  {journal} {\bibinfo
  {journal} {Europhysics Letters}\ }\textbf {\bibinfo {volume} {99}},\ \bibinfo
  {pages} {50004} (\bibinfo {year} {2012})}\BibitemShut {NoStop}%
\bibitem [{\citenamefont {Qi}\ \emph {et~al.}(2010)\citenamefont {Qi},
  \citenamefont {Hughes},\ and\ \citenamefont {Zhang}}]{Qi2010}%
  \BibitemOpen
  \bibfield  {author} {\bibinfo {author} {\bibfnamefont {X.-L.}\ \bibnamefont
  {Qi}}, \bibinfo {author} {\bibfnamefont {T.~L.}\ \bibnamefont {Hughes}},\
  and\ \bibinfo {author} {\bibfnamefont {S.-C.}\ \bibnamefont {Zhang}},\
  }\bibfield  {title} {\bibinfo {title} {Chiral topological superconductor from
  the quantum {H}all state},\ }\href
  {https://doi.org/10.1103/PhysRevB.82.184516} {\bibfield  {journal} {\bibinfo
  {journal} {Phys. Rev. B}\ }\textbf {\bibinfo {volume} {82}},\ \bibinfo
  {pages} {184516} (\bibinfo {year} {2010})}\BibitemShut {NoStop}%
\bibitem [{\citenamefont {Chang}\ \emph {et~al.}(2013)\citenamefont {Chang},
  \citenamefont {Zhang}, \citenamefont {Feng}, \citenamefont {Shen},
  \citenamefont {Zhang}, \citenamefont {Guo}, \citenamefont {Li}, \citenamefont
  {Ou}, \citenamefont {Wei}, \citenamefont {Wang}, \citenamefont {Ji},
  \citenamefont {Feng}, \citenamefont {Ji}, \citenamefont {Chen}, \citenamefont
  {Jia}, \citenamefont {Dai}, \citenamefont {Fang}, \citenamefont {Zhang},
  \citenamefont {He}, \citenamefont {Wang}, \citenamefont {Lu}, \citenamefont
  {Ma},\ and\ \citenamefont {Xue}}]{Chang2013}%
  \BibitemOpen
  \bibfield  {author} {\bibinfo {author} {\bibfnamefont {C.-Z.}\ \bibnamefont
  {Chang}}, \bibinfo {author} {\bibfnamefont {J.}~\bibnamefont {Zhang}},
  \bibinfo {author} {\bibfnamefont {X.}~\bibnamefont {Feng}}, \bibinfo {author}
  {\bibfnamefont {J.}~\bibnamefont {Shen}}, \bibinfo {author} {\bibfnamefont
  {Z.}~\bibnamefont {Zhang}}, \bibinfo {author} {\bibfnamefont
  {M.}~\bibnamefont {Guo}}, \bibinfo {author} {\bibfnamefont {K.}~\bibnamefont
  {Li}}, \bibinfo {author} {\bibfnamefont {Y.}~\bibnamefont {Ou}}, \bibinfo
  {author} {\bibfnamefont {P.}~\bibnamefont {Wei}}, \bibinfo {author}
  {\bibfnamefont {L.-L.}\ \bibnamefont {Wang}}, \bibinfo {author}
  {\bibfnamefont {Z.-Q.}\ \bibnamefont {Ji}}, \bibinfo {author} {\bibfnamefont
  {Y.}~\bibnamefont {Feng}}, \bibinfo {author} {\bibfnamefont {S.}~\bibnamefont
  {Ji}}, \bibinfo {author} {\bibfnamefont {X.}~\bibnamefont {Chen}}, \bibinfo
  {author} {\bibfnamefont {J.}~\bibnamefont {Jia}}, \bibinfo {author}
  {\bibfnamefont {X.}~\bibnamefont {Dai}}, \bibinfo {author} {\bibfnamefont
  {Z.}~\bibnamefont {Fang}}, \bibinfo {author} {\bibfnamefont {S.-C.}\
  \bibnamefont {Zhang}}, \bibinfo {author} {\bibfnamefont {K.}~\bibnamefont
  {He}}, \bibinfo {author} {\bibfnamefont {Y.}~\bibnamefont {Wang}}, \bibinfo
  {author} {\bibfnamefont {L.}~\bibnamefont {Lu}}, \bibinfo {author}
  {\bibfnamefont {X.-C.}\ \bibnamefont {Ma}},\ and\ \bibinfo {author}
  {\bibfnamefont {Q.-K.}\ \bibnamefont {Xue}},\ }\bibfield  {title} {\bibinfo
  {title} {Experimental observation of the quantum anomalous {H}all effect in a
  magnetic topological insulator},\ }\href
  {https://doi.org/10.1126/science.1234414} {\bibfield  {journal} {\bibinfo
  {journal} {Science}\ }\textbf {\bibinfo {volume} {340}},\ \bibinfo {pages}
  {167} (\bibinfo {year} {2013})}\BibitemShut {NoStop}%
\bibitem [{\citenamefont {Wang}\ \emph {et~al.}(2015)\citenamefont {Wang},
  \citenamefont {Zhou}, \citenamefont {Lian},\ and\ \citenamefont
  {Zhang}}]{Wang2015}%
  \BibitemOpen
  \bibfield  {author} {\bibinfo {author} {\bibfnamefont {J.}~\bibnamefont
  {Wang}}, \bibinfo {author} {\bibfnamefont {Q.}~\bibnamefont {Zhou}}, \bibinfo
  {author} {\bibfnamefont {B.}~\bibnamefont {Lian}},\ and\ \bibinfo {author}
  {\bibfnamefont {S.-C.}\ \bibnamefont {Zhang}},\ }\bibfield  {title} {\bibinfo
  {title} {Chiral topological superconductor and half-integer conductance
  plateau from quantum anomalous {H}all plateau transition},\ }\href
  {https://doi.org/10.1103/PhysRevB.92.064520} {\bibfield  {journal} {\bibinfo
  {journal} {Phys. Rev. B}\ }\textbf {\bibinfo {volume} {92}},\ \bibinfo
  {pages} {064520} (\bibinfo {year} {2015})}\BibitemShut {NoStop}%
\bibitem [{\citenamefont {Lian}\ \emph {et~al.}(2018)\citenamefont {Lian},
  \citenamefont {Sun}, \citenamefont {Vaezi}, \citenamefont {Qi},\ and\
  \citenamefont {Zhang}}]{Lian2018}%
  \BibitemOpen
  \bibfield  {author} {\bibinfo {author} {\bibfnamefont {B.}~\bibnamefont
  {Lian}}, \bibinfo {author} {\bibfnamefont {X.-Q.}\ \bibnamefont {Sun}},
  \bibinfo {author} {\bibfnamefont {A.}~\bibnamefont {Vaezi}}, \bibinfo
  {author} {\bibfnamefont {X.-L.}\ \bibnamefont {Qi}},\ and\ \bibinfo {author}
  {\bibfnamefont {S.-C.}\ \bibnamefont {Zhang}},\ }\bibfield  {title} {\bibinfo
  {title} {Topological quantum computation based on chiral {M}ajorana
  fermions},\ }\href {https://doi.org/10.1073/pnas.1810003115} {\bibfield
  {journal} {\bibinfo  {journal} {Proc. Natl. Acad. Sci.}\ }\textbf {\bibinfo
  {volume} {115}},\ \bibinfo {pages} {10938} (\bibinfo {year}
  {2018})}\BibitemShut {NoStop}%
\bibitem [{\citenamefont {Shen}\ \emph {et~al.}(2020)\citenamefont {Shen},
  \citenamefont {Lyu}, \citenamefont {Gao}, \citenamefont {Xie}, \citenamefont
  {Chen}, \citenamefont {woo Cho}, \citenamefont {Atanov}, \citenamefont
  {Chen}, \citenamefont {Liu}, \citenamefont {Hu}, \citenamefont {Yip},
  \citenamefont {Goh}, \citenamefont {He}, \citenamefont {Pan}, \citenamefont
  {Wang}, \citenamefont {Law},\ and\ \citenamefont {Lortz}}]{Shen}%
  \BibitemOpen
  \bibfield  {author} {\bibinfo {author} {\bibfnamefont {J.}~\bibnamefont
  {Shen}}, \bibinfo {author} {\bibfnamefont {J.}~\bibnamefont {Lyu}}, \bibinfo
  {author} {\bibfnamefont {J.~Z.}\ \bibnamefont {Gao}}, \bibinfo {author}
  {\bibfnamefont {Y.-M.}\ \bibnamefont {Xie}}, \bibinfo {author} {\bibfnamefont
  {C.-Z.}\ \bibnamefont {Chen}}, \bibinfo {author} {\bibfnamefont
  {C.}~\bibnamefont {woo Cho}}, \bibinfo {author} {\bibfnamefont
  {O.}~\bibnamefont {Atanov}}, \bibinfo {author} {\bibfnamefont
  {Z.}~\bibnamefont {Chen}}, \bibinfo {author} {\bibfnamefont {K.}~\bibnamefont
  {Liu}}, \bibinfo {author} {\bibfnamefont {Y.~J.}\ \bibnamefont {Hu}},
  \bibinfo {author} {\bibfnamefont {K.~Y.}\ \bibnamefont {Yip}}, \bibinfo
  {author} {\bibfnamefont {S.~K.}\ \bibnamefont {Goh}}, \bibinfo {author}
  {\bibfnamefont {Q.~L.}\ \bibnamefont {He}}, \bibinfo {author} {\bibfnamefont
  {L.}~\bibnamefont {Pan}}, \bibinfo {author} {\bibfnamefont {K.~L.}\
  \bibnamefont {Wang}}, \bibinfo {author} {\bibfnamefont {K.~T.}\ \bibnamefont
  {Law}},\ and\ \bibinfo {author} {\bibfnamefont {R.}~\bibnamefont {Lortz}},\
  }\bibfield  {title} {\bibinfo {title} {Spectroscopic fingerprint of chiral
  majorana modes at the edge of a quantum anomalous hall
  insulator/superconductor heterostructure},\ }\href
  {https://doi.org/10.1073/pnas.1910967117} {\bibfield  {journal} {\bibinfo
  {journal} {Proceedings of the National Academy of Sciences}\ }\textbf
  {\bibinfo {volume} {117}},\ \bibinfo {pages} {238} (\bibinfo {year}
  {2020})}\BibitemShut {NoStop}%
\bibitem [{\citenamefont {Kayyalha}\ \emph {et~al.}(2020)\citenamefont
  {Kayyalha}, \citenamefont {Xiao}, \citenamefont {Zhang}, \citenamefont
  {Shin}, \citenamefont {Jiang}, \citenamefont {Wang}, \citenamefont {Zhao},
  \citenamefont {Xiao}, \citenamefont {Zhang}, \citenamefont {Fijalkowski},
  \citenamefont {Mandal}, \citenamefont {Winnerlein}, \citenamefont {Gould},
  \citenamefont {Li}, \citenamefont {Molenkamp}, \citenamefont {Chan},
  \citenamefont {Samarth},\ and\ \citenamefont {Chang}}]{Kay2020}%
  \BibitemOpen
  \bibfield  {author} {\bibinfo {author} {\bibfnamefont {M.}~\bibnamefont
  {Kayyalha}}, \bibinfo {author} {\bibfnamefont {D.}~\bibnamefont {Xiao}},
  \bibinfo {author} {\bibfnamefont {R.}~\bibnamefont {Zhang}}, \bibinfo
  {author} {\bibfnamefont {J.}~\bibnamefont {Shin}}, \bibinfo {author}
  {\bibfnamefont {J.}~\bibnamefont {Jiang}}, \bibinfo {author} {\bibfnamefont
  {F.}~\bibnamefont {Wang}}, \bibinfo {author} {\bibfnamefont {Y.-F.}\
  \bibnamefont {Zhao}}, \bibinfo {author} {\bibfnamefont {R.}~\bibnamefont
  {Xiao}}, \bibinfo {author} {\bibfnamefont {L.}~\bibnamefont {Zhang}},
  \bibinfo {author} {\bibfnamefont {K.~M.}\ \bibnamefont {Fijalkowski}},
  \bibinfo {author} {\bibfnamefont {P.}~\bibnamefont {Mandal}}, \bibinfo
  {author} {\bibfnamefont {M.}~\bibnamefont {Winnerlein}}, \bibinfo {author}
  {\bibfnamefont {C.}~\bibnamefont {Gould}}, \bibinfo {author} {\bibfnamefont
  {Q.}~\bibnamefont {Li}}, \bibinfo {author} {\bibfnamefont {L.~W.}\
  \bibnamefont {Molenkamp}}, \bibinfo {author} {\bibfnamefont {M.~H.~W.}\
  \bibnamefont {Chan}}, \bibinfo {author} {\bibfnamefont {N.}~\bibnamefont
  {Samarth}},\ and\ \bibinfo {author} {\bibfnamefont {C.-Z.}\ \bibnamefont
  {Chang}},\ }\bibfield  {title} {\bibinfo {title} {Absence of evidence for
  chiral majorana modes in quantum anomalous hall-superconductor devices},\
  }\href {https://doi.org/10.1126/science.aax6361} {\bibfield  {journal}
  {\bibinfo  {journal} {Science}\ }\textbf {\bibinfo {volume} {367}},\ \bibinfo
  {pages} {64} (\bibinfo {year} {2020})}\BibitemShut {NoStop}%
\bibitem [{\citenamefont {Uday}\ \emph {et~al.}(2025)\citenamefont {Uday},
  \citenamefont {Lippertz}, \citenamefont {Bhujel}, \citenamefont {Taskin},\
  and\ \citenamefont {Ando}}]{Uday2025}%
  \BibitemOpen
  \bibfield  {author} {\bibinfo {author} {\bibfnamefont {A.}~\bibnamefont
  {Uday}}, \bibinfo {author} {\bibfnamefont {G.}~\bibnamefont {Lippertz}},
  \bibinfo {author} {\bibfnamefont {B.}~\bibnamefont {Bhujel}}, \bibinfo
  {author} {\bibfnamefont {A.~A.}\ \bibnamefont {Taskin}},\ and\ \bibinfo
  {author} {\bibfnamefont {Y.}~\bibnamefont {Ando}},\ }\bibfield  {title}
  {\bibinfo {title} {Non-{M}ajorana origin of the half-integer conductance
  quantization elucidated by multiterminal superconductor--quantum anomalous
  {H}all insulator heterostructure},\ }\href
  {https://doi.org/10.1103/PhysRevB.111.035440} {\bibfield  {journal} {\bibinfo
   {journal} {Phys. Rev. B}\ }\textbf {\bibinfo {volume} {111}},\ \bibinfo
  {pages} {035440} (\bibinfo {year} {2025})}\BibitemShut {NoStop}%
\bibitem [{\citenamefont {Ji}\ and\ \citenamefont {Wen}(2018)}]{Ji2018}%
  \BibitemOpen
  \bibfield  {author} {\bibinfo {author} {\bibfnamefont {W.}~\bibnamefont
  {Ji}}\ and\ \bibinfo {author} {\bibfnamefont {X.-G.}\ \bibnamefont {Wen}},\
  }\bibfield  {title} {\bibinfo {title} {$\frac{1}{2}({e}^{2}/h)$ conductance
  plateau without 1d chiral {M}ajorana fermions},\ }\href
  {https://doi.org/10.1103/PhysRevLett.120.107002} {\bibfield  {journal}
  {\bibinfo  {journal} {Phys. Rev. Lett.}\ }\textbf {\bibinfo {volume} {120}},\
  \bibinfo {pages} {107002} (\bibinfo {year} {2018})}\BibitemShut {NoStop}%
\bibitem [{\citenamefont {Huang}\ \emph {et~al.}(2018)\citenamefont {Huang},
  \citenamefont {Setiawan},\ and\ \citenamefont {Sau}}]{Huang2018}%
  \BibitemOpen
  \bibfield  {author} {\bibinfo {author} {\bibfnamefont {Y.}~\bibnamefont
  {Huang}}, \bibinfo {author} {\bibfnamefont {F.}~\bibnamefont {Setiawan}},\
  and\ \bibinfo {author} {\bibfnamefont {J.~D.}\ \bibnamefont {Sau}},\
  }\bibfield  {title} {\bibinfo {title} {Disorder-induced half-integer
  quantized conductance plateau in quantum anomalous {H}all
  insulator-superconductor structures},\ }\href
  {https://doi.org/10.1103/PhysRevB.97.100501} {\bibfield  {journal} {\bibinfo
  {journal} {Phys. Rev. B}\ }\textbf {\bibinfo {volume} {97}},\ \bibinfo
  {pages} {100501} (\bibinfo {year} {2018})}\BibitemShut {NoStop}%
\bibitem [{\citenamefont {Klees}\ \emph {et~al.}(2024)\citenamefont {Klees},
  \citenamefont {Gresta}, \citenamefont {Sturm}, \citenamefont {Molenkamp},\
  and\ \citenamefont {Hankiewicz}}]{Klees2024}%
  \BibitemOpen
  \bibfield  {author} {\bibinfo {author} {\bibfnamefont {R.~L.}\ \bibnamefont
  {Klees}}, \bibinfo {author} {\bibfnamefont {D.}~\bibnamefont {Gresta}},
  \bibinfo {author} {\bibfnamefont {J.}~\bibnamefont {Sturm}}, \bibinfo
  {author} {\bibfnamefont {L.~W.}\ \bibnamefont {Molenkamp}},\ and\ \bibinfo
  {author} {\bibfnamefont {E.~M.}\ \bibnamefont {Hankiewicz}},\ }\bibfield
  {title} {\bibinfo {title} {Majorana-mediated thermoelectric transport in
  multiterminal junctions},\ }\href
  {https://doi.org/10.1103/PhysRevB.110.064517} {\bibfield  {journal} {\bibinfo
   {journal} {Phys. Rev. B}\ }\textbf {\bibinfo {volume} {110}},\ \bibinfo
  {pages} {064517} (\bibinfo {year} {2024})}\BibitemShut {NoStop}%
\bibitem [{\citenamefont {Osca}\ and\ \citenamefont {Serra}(2018)}]{Osca2018}%
  \BibitemOpen
  \bibfield  {author} {\bibinfo {author} {\bibfnamefont {J.}~\bibnamefont
  {Osca}}\ and\ \bibinfo {author} {\bibfnamefont {L.}~\bibnamefont {Serra}},\
  }\bibfield  {title} {\bibinfo {title} {Conductance oscillations and speed of
  chiral {M}ajorana mode in a quantum-anomalous-{H}all 2d strip},\ }\href
  {https://doi.org/10.1103/PhysRevB.98.121407} {\bibfield  {journal} {\bibinfo
  {journal} {Phys. Rev. B}\ }\textbf {\bibinfo {volume} {98}},\ \bibinfo
  {pages} {121407} (\bibinfo {year} {2018})}\BibitemShut {NoStop}%
\bibitem [{\citenamefont {Osca}\ and\ \citenamefont {Serra}(2025)}]{Osca2026}%
  \BibitemOpen
  \bibfield  {author} {\bibinfo {author} {\bibfnamefont {J.}~\bibnamefont
  {Osca}}\ and\ \bibinfo {author} {\bibfnamefont {L.}~\bibnamefont {Serra}},\
  }\bibfield  {title} {\bibinfo {title} {Electrostatic gating and the
  interference of chiral majoranas in thin slabs of magnetic topological
  insulators},\ }\href {https://doi.org/10.1103/dbd6-8h9b} {\bibfield
  {journal} {\bibinfo  {journal} {Phys. Rev. B}\ }\textbf {\bibinfo {volume}
  {112}},\ \bibinfo {pages} {224519} (\bibinfo {year} {2025})}\BibitemShut
  {NoStop}%
\bibitem [{\citenamefont {Barone}\ and\ \citenamefont
  {Patern{\`o}}(1982)}]{BaPa1982}%
  \BibitemOpen
  \bibfield  {author} {\bibinfo {author} {\bibfnamefont {A.}~\bibnamefont
  {Barone}}\ and\ \bibinfo {author} {\bibfnamefont {G.}~\bibnamefont
  {Patern{\`o}}},\ }\href@noop {} {\emph {\bibinfo {title} {Physics and
  Applications of the Josephson Effect}}}\ (\bibinfo  {publisher} {Wiley},\
  \bibinfo {year} {1982})\BibitemShut {NoStop}%
\bibitem [{\citenamefont {Tinkham}(2004)}]{Tink2004}%
  \BibitemOpen
  \bibfield  {author} {\bibinfo {author} {\bibfnamefont {M.}~\bibnamefont
  {Tinkham}},\ }\href@noop {} {\emph {\bibinfo {title} {Introduction to
  Superconductivity}}},\ \bibinfo {edition} {2nd}\ ed.\ (\bibinfo  {publisher}
  {Dover},\ \bibinfo {year} {2004})\BibitemShut {NoStop}%
\bibitem [{\citenamefont {Golubov}\ \emph {et~al.}(2004)\citenamefont
  {Golubov}, \citenamefont {Kupriyanov},\ and\ \citenamefont
  {Il'ichev}}]{Golu2004}%
  \BibitemOpen
  \bibfield  {author} {\bibinfo {author} {\bibfnamefont {A.~A.}\ \bibnamefont
  {Golubov}}, \bibinfo {author} {\bibfnamefont {M.~Y.}\ \bibnamefont
  {Kupriyanov}},\ and\ \bibinfo {author} {\bibfnamefont {E.}~\bibnamefont
  {Il'ichev}},\ }\bibfield  {title} {\bibinfo {title} {The current-phase
  relation in josephson junctions},\ }\href
  {https://doi.org/10.1103/RevModPhys.76.411} {\bibfield  {journal} {\bibinfo
  {journal} {Rev. Mod. Phys.}\ }\textbf {\bibinfo {volume} {76}},\ \bibinfo
  {pages} {411} (\bibinfo {year} {2004})}\BibitemShut {NoStop}%
\bibitem [{\citenamefont {Antonelli}\ \emph {et~al.}(2025)\citenamefont
  {Antonelli}, \citenamefont {Coraiola}, \citenamefont {Ohnmacht},
  \citenamefont {Svetogorov}, \citenamefont {Sabonis}, \citenamefont {ten
  Kate}, \citenamefont {Cheah}, \citenamefont {Krizek}, \citenamefont {Schott},
  \citenamefont {Cuevas}, \citenamefont {Belzig}, \citenamefont {Wegscheider},\
  and\ \citenamefont {Nichele}}]{Anto2025}%
  \BibitemOpen
  \bibfield  {author} {\bibinfo {author} {\bibfnamefont {T.}~\bibnamefont
  {Antonelli}}, \bibinfo {author} {\bibfnamefont {M.}~\bibnamefont {Coraiola}},
  \bibinfo {author} {\bibfnamefont {D.~C.}\ \bibnamefont {Ohnmacht}}, \bibinfo
  {author} {\bibfnamefont {A.~E.}\ \bibnamefont {Svetogorov}}, \bibinfo
  {author} {\bibfnamefont {D.}~\bibnamefont {Sabonis}}, \bibinfo {author}
  {\bibfnamefont {S.~C.}\ \bibnamefont {ten Kate}}, \bibinfo {author}
  {\bibfnamefont {E.}~\bibnamefont {Cheah}}, \bibinfo {author} {\bibfnamefont
  {F.}~\bibnamefont {Krizek}}, \bibinfo {author} {\bibfnamefont
  {R.}~\bibnamefont {Schott}}, \bibinfo {author} {\bibfnamefont {J.~C.}\
  \bibnamefont {Cuevas}}, \bibinfo {author} {\bibfnamefont {W.}~\bibnamefont
  {Belzig}}, \bibinfo {author} {\bibfnamefont {W.}~\bibnamefont
  {Wegscheider}},\ and\ \bibinfo {author} {\bibfnamefont {F.}~\bibnamefont
  {Nichele}},\ }\bibfield  {title} {\bibinfo {title} {Exploring the energy
  spectrum of a four-terminal josephson junction: Toward topological andreev
  band structures},\ }\href {https://doi.org/10.1103/qd3y-f912} {\bibfield
  {journal} {\bibinfo  {journal} {Phys. Rev. X}\ }\textbf {\bibinfo {volume}
  {15}},\ \bibinfo {pages} {031066} (\bibinfo {year} {2025})}\BibitemShut
  {NoStop}%
\bibitem [{\citenamefont {Zazunov}\ \emph {et~al.}(2017)\citenamefont
  {Zazunov}, \citenamefont {Egger}, \citenamefont {Alvarado},\ and\
  \citenamefont {Yeyati}}]{Zazu2017}%
  \BibitemOpen
  \bibfield  {author} {\bibinfo {author} {\bibfnamefont {A.}~\bibnamefont
  {Zazunov}}, \bibinfo {author} {\bibfnamefont {R.}~\bibnamefont {Egger}},
  \bibinfo {author} {\bibfnamefont {M.}~\bibnamefont {Alvarado}},\ and\
  \bibinfo {author} {\bibfnamefont {A.~L.}\ \bibnamefont {Yeyati}},\ }\bibfield
   {title} {\bibinfo {title} {Josephson effect in multiterminal topological
  junctions},\ }\href {https://doi.org/10.1103/PhysRevB.96.024516} {\bibfield
  {journal} {\bibinfo  {journal} {Phys. Rev. B}\ }\textbf {\bibinfo {volume}
  {96}},\ \bibinfo {pages} {024516} (\bibinfo {year} {2017})}\BibitemShut
  {NoStop}%
\bibitem [{\citenamefont {Schiela}\ \emph {et~al.}(2024)\citenamefont
  {Schiela}, \citenamefont {Yu},\ and\ \citenamefont {Shabani}}]{Schiela2024}%
  \BibitemOpen
  \bibfield  {author} {\bibinfo {author} {\bibfnamefont {W.~F.}\ \bibnamefont
  {Schiela}}, \bibinfo {author} {\bibfnamefont {P.}~\bibnamefont {Yu}},\ and\
  \bibinfo {author} {\bibfnamefont {J.}~\bibnamefont {Shabani}},\ }\bibfield
  {title} {\bibinfo {title} {Progress in superconductor-semiconductor
  topological josephson junctions},\ }\href
  {https://doi.org/10.1103/PRXQuantum.5.030102} {\bibfield  {journal} {\bibinfo
   {journal} {PRX Quantum}\ }\textbf {\bibinfo {volume} {5}},\ \bibinfo {pages}
  {030102} (\bibinfo {year} {2024})}\BibitemShut {NoStop}%
\bibitem [{\citenamefont {Di~Miceli}\ \emph {et~al.}(2023)\citenamefont
  {Di~Miceli}, \citenamefont {Zsurka}, \citenamefont {Legendre}, \citenamefont
  {Moors}, \citenamefont {Schmidt},\ and\ \citenamefont {Serra}}]{Daniele}%
  \BibitemOpen
  \bibfield  {author} {\bibinfo {author} {\bibfnamefont {D.}~\bibnamefont
  {Di~Miceli}}, \bibinfo {author} {\bibfnamefont {E.}~\bibnamefont {Zsurka}},
  \bibinfo {author} {\bibfnamefont {J.}~\bibnamefont {Legendre}}, \bibinfo
  {author} {\bibfnamefont {K.}~\bibnamefont {Moors}}, \bibinfo {author}
  {\bibfnamefont {T.~L.}\ \bibnamefont {Schmidt}},\ and\ \bibinfo {author}
  {\bibfnamefont {L.}~\bibnamefont {Serra}},\ }\bibfield  {title} {\bibinfo
  {title} {Conductance asymmetry in proximitized magnetic topological insulator
  junctions with majorana modes},\ }\href
  {https://doi.org/10.1103/PhysRevB.108.035424} {\bibfield  {journal} {\bibinfo
   {journal} {Phys. Rev. B}\ }\textbf {\bibinfo {volume} {108}},\ \bibinfo
  {pages} {035424} (\bibinfo {year} {2023})}\BibitemShut {NoStop}%
\bibitem [{\citenamefont {Datta}(1995)}]{Datta}%
  \BibitemOpen
  \bibfield  {author} {\bibinfo {author} {\bibfnamefont {S.}~\bibnamefont
  {Datta}},\ }\href {https://doi.org/10.1017/CBO9780511805776} {\emph {\bibinfo
  {title} {Electronic Transport in Mesoscopic Systems}}},\ Cambridge Studies in
  Semiconductor Physics and Microelectronic Engineering\ (\bibinfo  {publisher}
  {Cambridge University Press},\ \bibinfo {address} {Cambridge},\ \bibinfo
  {year} {1995})\BibitemShut {NoStop}%
\bibitem [{\citenamefont {Cahay}\ \emph {et~al.}(1988)\citenamefont {Cahay},
  \citenamefont {McLennan},\ and\ \citenamefont {Datta}}]{Cahay1988}%
  \BibitemOpen
  \bibfield  {author} {\bibinfo {author} {\bibfnamefont {M.}~\bibnamefont
  {Cahay}}, \bibinfo {author} {\bibfnamefont {M.}~\bibnamefont {McLennan}},\
  and\ \bibinfo {author} {\bibfnamefont {S.}~\bibnamefont {Datta}},\ }\bibfield
   {title} {\bibinfo {title} {Conductance of an array of elastic scatterers: A
  scattering-matrix approach},\ }\href
  {https://doi.org/10.1103/PhysRevB.37.10125} {\bibfield  {journal} {\bibinfo
  {journal} {Phys. Rev. B}\ }\textbf {\bibinfo {volume} {37}},\ \bibinfo
  {pages} {10125} (\bibinfo {year} {1988})}\BibitemShut {NoStop}%
\bibitem [{\citenamefont {Serra}\ and\ \citenamefont {Choi}(2009)}]{Serra2009}%
  \BibitemOpen
  \bibfield  {author} {\bibinfo {author} {\bibfnamefont {L.}~\bibnamefont
  {Serra}}\ and\ \bibinfo {author} {\bibfnamefont {M.-S.}\ \bibnamefont
  {Choi}},\ }\bibfield  {title} {\bibinfo {title} {Conductance of tubular
  nanowires with disorder},\ }\href
  {https://doi.org/10.1140/epjb/e2009-00280-6} {\bibfield  {journal} {\bibinfo
  {journal} {The European Physical Journal B}\ }\textbf {\bibinfo {volume}
  {71}},\ \bibinfo {pages} {97} (\bibinfo {year} {2009})}\BibitemShut {NoStop}%
\bibitem [{\citenamefont {Estarellas}\ and\ \citenamefont
  {Serra}(2016)}]{Estarellas}%
  \BibitemOpen
  \bibfield  {author} {\bibinfo {author} {\bibfnamefont {C.}~\bibnamefont
  {Estarellas}}\ and\ \bibinfo {author} {\bibfnamefont {L.}~\bibnamefont
  {Serra}},\ }\bibfield  {title} {\bibinfo {title} {Resonant anderson
  localization in segmented wires},\ }\href
  {https://doi.org/10.1103/PhysRevE.93.032105} {\bibfield  {journal} {\bibinfo
  {journal} {Phys. Rev. E}\ }\textbf {\bibinfo {volume} {93}},\ \bibinfo
  {pages} {032105} (\bibinfo {year} {2016})}\BibitemShut {NoStop}%
\bibitem [{\citenamefont {Osca}\ and\ \citenamefont {Serra}(2019)}]{Osca2019}%
  \BibitemOpen
  \bibfield  {author} {\bibinfo {author} {\bibfnamefont {J.}~\bibnamefont
  {Osca}}\ and\ \bibinfo {author} {\bibfnamefont {L.}~\bibnamefont {Serra}},\
  }\bibfield  {title} {\bibinfo {title} {Complex band-structure analysis and
  topological physics of majorana nanowires},\ }\href
  {https://doi.org/10.1140/epjb/e2019-100011-2} {\bibfield  {journal} {\bibinfo
   {journal} {The European Physical Journal B}\ }\textbf {\bibinfo {volume}
  {92}},\ \bibinfo {pages} {101} (\bibinfo {year} {2019})}\BibitemShut
  {NoStop}%
\bibitem [{\citenamefont {Taddei}\ and\ \citenamefont {Fazio}(2023)}]{Taddei}%
  \BibitemOpen
  \bibfield  {author} {\bibinfo {author} {\bibfnamefont {F.}~\bibnamefont
  {Taddei}}\ and\ \bibinfo {author} {\bibfnamefont {R.}~\bibnamefont {Fazio}},\
  }\bibfield  {title} {\bibinfo {title} {Thermodynamic uncertainty relations
  for systems with broken time reversal symmetry: The case of superconducting
  hybrid systems},\ }\href {https://doi.org/10.1103/PhysRevB.108.115422}
  {\bibfield  {journal} {\bibinfo  {journal} {Phys. Rev. B}\ }\textbf {\bibinfo
  {volume} {108}},\ \bibinfo {pages} {115422} (\bibinfo {year}
  {2023})}\BibitemShut {NoStop}%
\bibitem [{\citenamefont {Mateos}\ \emph {et~al.}(2024)\citenamefont {Mateos},
  \citenamefont {Tosi}, \citenamefont {Braggio}, \citenamefont {Taddei},\ and\
  \citenamefont {Arrachea}}]{Liliana}%
  \BibitemOpen
  \bibfield  {author} {\bibinfo {author} {\bibfnamefont {J.~H.}\ \bibnamefont
  {Mateos}}, \bibinfo {author} {\bibfnamefont {L.}~\bibnamefont {Tosi}},
  \bibinfo {author} {\bibfnamefont {A.}~\bibnamefont {Braggio}}, \bibinfo
  {author} {\bibfnamefont {F.}~\bibnamefont {Taddei}},\ and\ \bibinfo {author}
  {\bibfnamefont {L.}~\bibnamefont {Arrachea}},\ }\bibfield  {title} {\bibinfo
  {title} {Nonlocal thermoelectricity in quantum wires as a signature of
  bogoliubov-fermi points},\ }\href
  {https://doi.org/10.1103/PhysRevB.110.075415} {\bibfield  {journal} {\bibinfo
   {journal} {Phys. Rev. B}\ }\textbf {\bibinfo {volume} {110}},\ \bibinfo
  {pages} {075415} (\bibinfo {year} {2024})}\BibitemShut {NoStop}%
\bibitem [{\citenamefont {Arrachea}\ \emph {et~al.}(2025)\citenamefont
  {Arrachea}, \citenamefont {Braggio}, \citenamefont {Burset}, \citenamefont
  {Lee}, \citenamefont {Levy~Yeyati},\ and\ \citenamefont
  {S{\'a}nchez}}]{Arra2025}%
  \BibitemOpen
  \bibfield  {author} {\bibinfo {author} {\bibfnamefont {L.}~\bibnamefont
  {Arrachea}}, \bibinfo {author} {\bibfnamefont {A.}~\bibnamefont {Braggio}},
  \bibinfo {author} {\bibfnamefont {P.}~\bibnamefont {Burset}}, \bibinfo
  {author} {\bibfnamefont {E.~J.~H.}\ \bibnamefont {Lee}}, \bibinfo {author}
  {\bibfnamefont {A.}~\bibnamefont {Levy~Yeyati}},\ and\ \bibinfo {author}
  {\bibfnamefont {R.}~\bibnamefont {S{\'a}nchez}},\ }\bibfield  {title}
  {\bibinfo {title} {Thermoelectric processes of quantum normal-superconductor
  interfaces},\ }\href {https://doi.org/https://doi.org/10.1002/andp.202500197}
  {\bibfield  {journal} {\bibinfo  {journal} {Annalen der Physik}\ }\textbf
  {\bibinfo {volume} {537}},\ \bibinfo {pages} {e00197} (\bibinfo {year}
  {2025})}\BibitemShut {NoStop}%
\bibitem [{\citenamefont {Ashcroft}\ and\ \citenamefont
  {Mermin}(1976)}]{Ashcroft76}%
  \BibitemOpen
  \bibfield  {author} {\bibinfo {author} {\bibfnamefont {N.~W.}\ \bibnamefont
  {Ashcroft}}\ and\ \bibinfo {author} {\bibfnamefont {N.~D.}\ \bibnamefont
  {Mermin}},\ }\href@noop {} {\emph {\bibinfo {title} {{S}olid {S}tate
  {P}hysics}}}\ (\bibinfo  {publisher} {Holt-Saunders},\ \bibinfo {year}
  {1976})\BibitemShut {NoStop}%
\end{thebibliography}%

\appendix
\section{Analytic composition of scattering matrices}
\label{appA}

We write the scattering matrix of a single superconducting island (NSN junction) as
\begin{equation}
S(\xi,\theta) =
\begin{pmatrix}
r_{ee} & r_{he}  & t'_{ee} & t'_{he}\\
r_{eh} & r_{hh}  & t'_{eh} & t'_{hh}\\
t_{ee} & t_{he}  & r'_{ee}  & r'_{he} \\
t_{eh} & t_{hh}  & r'_{eh}  & r'_{hh}
\end{pmatrix}.
\end{equation}

For two sequential islands forming an NSNSN device [Fig.~\ref{F0B}b], the outgoing modes from junction 1 that propagate into the central normal region match the left-incident modes of junction 2, up to a phase factor, and vice versa. Therefore,
\begin{align}
a_{re1} &= b_{le2}\, e^{i\mathcal{L}}, \nonumber\\
a_{rh1} &= b_{lh2}\, e^{i\mathcal{L}}, \nonumber\\
a_{le2} &= b_{re1}\, e^{i\mathcal{L}}, \nonumber\\
a_{lh2} &= b_{rh1}\, e^{i\mathcal{L}}.
\label{eqA2}
\end{align}

Assuming that ingoing modes originate in the left lead only
and using Eq.~(\ref{eqA2})
as well as
the definitions $b_1=S a_1$, $b_2=S a_2$,
it is possible    
to derive after some algebra
\begin{widetext}
\begin{align}
b_{le1} &= 
\frac{1}{2} + \frac{e^{2 i (\xi + \mathcal{L})} 
\left(-1 + \cos\phi_{12}\right)}{-4 + 2 e^{2 i \mathcal{L}} 
\left(1 + \cos\phi_{12}\right) }\; , \\[0.8em]
b_{re1} &= \frac{e^{i (\xi + \phi_2)} \left(e^{i (2 \mathcal{L} + \phi_2)} + e^{i \phi_1} \left(-2 + e^{2 i \mathcal{L}}\right)\right)}{e^{2 i (\mathcal{L} + \phi_1)} + e^{2 i (\mathcal{L} + \phi_2)} - 4 e^{i \phi_{12}} + 2 e^{i (2 \mathcal{L} + \phi_{12})}}\; , \\[0.8em]
b_{lh1} & = 
\frac{e^{-i \phi_1} \Big(\cos\mathcal{L} - 3 i \sin\mathcal{L} - (\cos\xi + i (1 - 2 
\cos\phi_{12}
)) 
\sin\xi \, (\cos(\xi + \mathcal{L}) + i \sin(\xi + \mathcal{L}))\Big)}{2 \Big(-\cos\mathcal{L} + e^{i \mathcal{L}} 
\cos\phi_{12} 
+ 3 i \sin\mathcal{L}\Big)}\; , \\[0.8em]
b_{rh1} & = \frac{e^{i \xi} \left(e^{i (2 \mathcal{L} + \phi_1)} + e^{i \phi_2} \left(-2 + e^{2 i \mathcal{L}}\right)\right)}{e^{2 i (\mathcal{L} + \phi_1)} + e^{2 i (\mathcal{L} + \phi_2)} + 2 
e^{i \phi_{12}} 
\left(-2 + e^{2 i \mathcal{L}}\right)}\; , \\[0.8em]
b_{le2} & = \frac{e^{i (\xi + \mathcal{L} + \phi_2)} \left(- e^{i \phi_1} + e^{i \phi_2}\right)}{e^{2 i (\mathcal{L} + \phi_1)} + e^{2 i (\mathcal{L} + \phi_2)} + 2 
e^{i \phi_{12}} 
\left(-2 + e^{2 i \mathcal{L}}\right)}\; , \\[0.8em]
b_{re2} & = \frac{e^{i (2 \xi + \mathcal{L} + \phi_2)} \left(-1 + e^{2 i \mathcal{L}}\right) \left(e^{i \phi_1} + e^{i \phi_2}\right)}{e^{2 i (\mathcal{L} + \phi_1)} + e^{2 i (\mathcal{L} + \phi_2)} + 2 
e^{i \phi_{12}} 
\left(-2 + e^{2 i \mathcal{L}}\right)}\,, \\[0.8em]
b_{lh2} & = \frac{e^{i (\xi + \mathcal{L})} \left(e^{i \phi_1} - e^{i \phi_2}\right)}{e^{2 i (\mathcal{L} + \phi_1)} + e^{2 i (\mathcal{L} + \phi_2)} + 2 
e^{i \phi_{12}} 
\left(-2 + e^{2 i \mathcal{L}}\right)}\; , \\[0.8em]
b_{rh2} &= \frac{e^{i (2 \xi + \mathcal{L})} \left(-1 + e^{2 i \mathcal{L}}\right) \left(e^{i \phi_1} + e^{i \phi_2}\right)}{e^{2 i (\mathcal{L} + \phi_1)} + e^{2 i (\mathcal{L} + \phi_2)} + 2 e^{i (\phi_1 + \phi_2)} \left(-2 + e^{2 i \mathcal{L}}\right)}\; .
\end{align}
\end{widetext}

The extension to the three-terminal system [Fig.~\ref{F10}] follows a similar, albeit more cumbersome, procedure. In this case, one must account for the phases accumulated along the three segments $\ell_{1,2,3}$ separating the superconducting islands.

\section{Numerical modeling}
\label{appB}

\begin{figure}[b]
  \centering
  \includegraphics[width=0.4\textwidth]{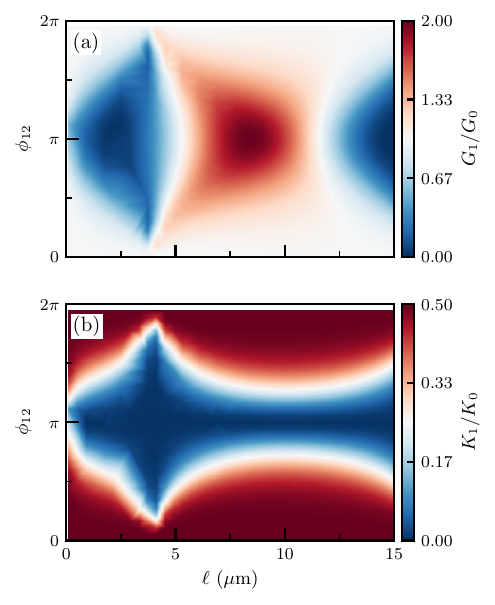}
\caption{
(a) Conductance of a two-terminal device, calculated numerically as a function of $\phi_{12}$ and $\ell$. 
(b) The same for the thermal conductance. 
The remaining parameters are $L_y = 2.5\,\mu\mathrm{m}$, $L_s = 23\,\mu\mathrm{m}$, $\alpha = 0.2\,\mathrm{meV}\,\mu\mathrm{m}$, 
$m_0 = 17\,\mathrm{meV}$, and $\hbar m_1 = 10^{-3}\,\mathrm{meV}\,\mu\mathrm{m}^2$. 
Additionally, $\Delta_t = 1\,\mathrm{meV}$ and $\Delta_b = 0$ in the superconducting regions, 
while $\Delta_t = \Delta_b = 0$ in the normal regions.
}  
\label{F5}
\end{figure}

The probability and charge densities shown in Fig.~\ref{F5} are obtained using a numerical method based on complex band structure analysis (see Ref.~\cite{Osca2026} for a review in the context of Majorana nanowires). In each homogeneous region of the junction, the wave function is expressed as a linear superposition of plane-wave modes,
\begin{equation}
\label{eq2A}
\Psi (x,y) = \sum_{k,\sigma,\tau,\lambda} 
C_k\,
\Phi^{(k)}_{\sigma\tau\lambda}(y)\,
e^{i k x},
\end{equation}
where $k$ spans a set of complex wave numbers compatible with a given energy $E$. The coefficients $C_k$ are the corresponding amplitudes in each region, while $\Phi^{(k)}_{\sigma\tau\lambda}(y)$ are eight-component spinors ($\sigma,\tau,\lambda = 1,2$).
Importantly, this method remains valid even for short superconducting segments $L_s$, where tunneling effects become significant.

The numerical algorithm proceeds in two steps. First, the Hamiltonian of each homogeneous region is diagonalized assuming infinite length, yielding the complex wave numbers $k$ and the corresponding eigenmodes at energy $E$,
\begin{equation}
\label{eqA1}
\mathcal{H}_k\, \Phi^{(k)}(y) = E\, \Phi^{(k)}(y)\;,
\end{equation}
where $\mathcal{H}_k$ is obtained from Eq.~(\ref{EQ1}) by replacing $p_x \to \hbar k$.
Second, the set of amplitudes $\{C_k\}$ in each region is determined by solving a linear system that enforces the matching conditions at the interfaces. The resulting amplitudes, together with the eigenmodes $\Phi^{(k)}$, fully determine the scattering-state wave function $\Psi$, from which the probability and charge densities are directly obtained.

\begin{figure}[b]
  \centering
  \includegraphics[width=0.4\textwidth]{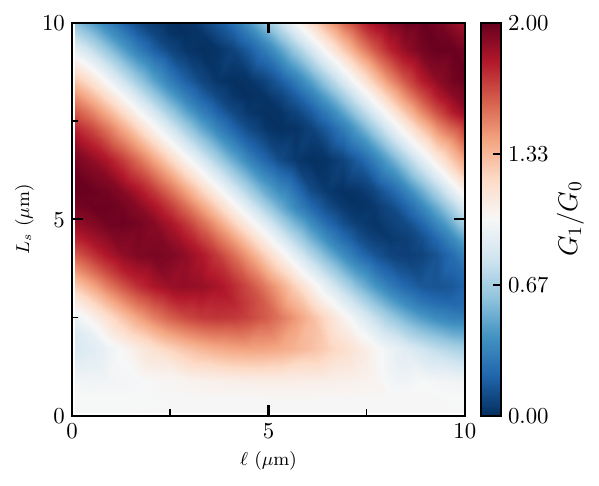}
\caption{
Conductance of a two-terminal device calculated numerically as a function of $\ell$ and $L_s$. 
The remaining parameters are $L_y = 2.5\,\mu\mathrm{m}$, $L_s = 23\,\mu\mathrm{m}$, 
$\alpha = 0.2\,\mathrm{meV}\,\mu\mathrm{m}$, $m_0 = 17\,\mathrm{meV}$, and 
$\hbar m_1 = 10^{-3}\,\mathrm{meV}\,\mu\mathrm{m}^2$. 
Additionally, $\Delta_t = 1\,\mathrm{meV}$ and $\Delta_b = 0$ in the superconducting regions, 
while $\Delta_t = \Delta_b = 0$ in the normal regions.
}
\label{F6}
\end{figure}

Figures~\ref{F5} and \ref{F6} show the electrical and thermal conductances obtained from Eqs.~(\ref{EQ11}) and (\ref{EQ12}), using numerically calculated probabilities in place of the analytical ones. 
In Fig.~\ref{F5}, these quantities are plotted as functions of the superconducting phase difference $\phi_{12}$ and the separation between superconductors $\ell$. The results closely reproduce those of Fig.~\ref{F2}, confirming the analytical predictions.
Similarly, Fig.~\ref{F6} shows the conductance $G_1/G_0$ as a function of the superconducting segment length $L_s$ and the separation $\ell$, corresponding to the analytical results in Fig.~\ref{F4}. 
Taken together, these figures provide numerical validation of the analytical findings.

\begin{figure}[p]
  \centering
  \includegraphics[width=0.4\textwidth]{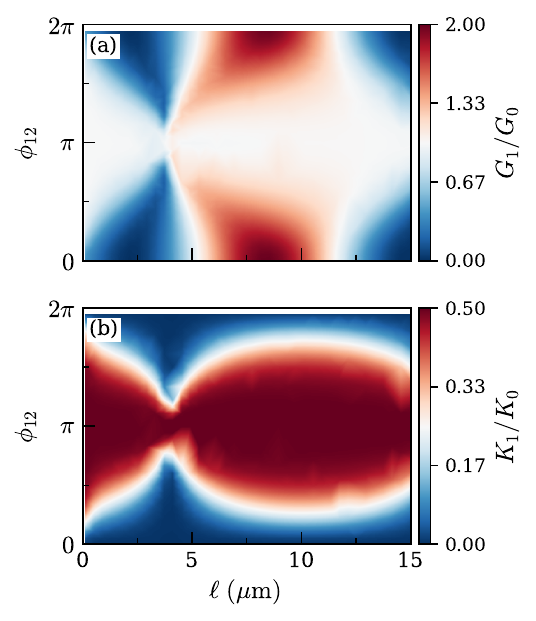}
\caption{
(a) Conductance of a two-terminal device in which the superconductors are proximitized to opposite sides of the MTI. 
The calculation is performed numerically as a function of $\phi_{12}$ and $\ell$. 
(b) The same for the thermal conductance. 
The remaining parameters are $L_y = 2.5\,\mu\mathrm{m}$, $L_s = 23\,\mu\mathrm{m}$, 
$\alpha = 0.2\,\mathrm{meV}\,\mu\mathrm{m}$, $m_0 = 17\,\mathrm{meV}$, and 
$\hbar m_1 = 10^{-3}\,\mathrm{meV}\,\mu\mathrm{m}^2$. 
}
\label{F12}
\end{figure}

Finally, Fig.~\ref{F12} shows the electrical and thermal conductances for the case in which the right superconducting region is proximity coupled to an external superconductor in contact with the bottom surface of the MTI slab, while the left superconducting region remains coupled to a superconductor on the top surface.
This asymmetric proximity configuration induces a phase shift of $\pi$, which displaces the blockade region along the vertical axis relative to the case where both superconductors are coupled from the same side. Consequently, the blockade condition is modified to $\phi_{12} = 0,\, 2\pi$.

\end{document}